\documentclass[%
 reprint,
superscriptaddress,
showpacs,preprintnumbers,
 amsmath,amssymb,
 aps,
pra,
]{revtex4-1}
\pdfoutput=1
\usepackage{slashed}
\usepackage{graphicx}
\usepackage{dcolumn}
\usepackage{bm}

\begin{document}

\title{Wave functions for high-symmetry, thin  microstrip antennas and two-dimensional quantum boxes}



\author{Joseph R. Rain}
\affiliation{Department of Physics, University of Central Florida, 4111 Libra Drive, Orlando, Florida, 32816-2385, USA}
\author{PeiYu Cai}
\affiliation{Department of Physics, University of Central Florida, 4111 Libra Drive, Orlando, Florida, 32816-2385, USA}
\affiliation{University College London, Department of Physics, South Kensington Campus, London SW7 2AZ, UK*}
\author{Alexander Baekey}
\affiliation{Department of Physics, University of Central Florida, 4111 Libra Drive, Orlando, Florida, 32816-2385, USA}

\author{Matthew A. Reinhard}
\affiliation{Department of Physics, University of Central Florida, 4111 Libra Drive, Orlando, Florida, 32816-2385, USA}
\affiliation{Department of Physics, University of Florida, 2001 Museum Road, P. O. Box 118440, Gainesville, Florida, 32611-8440, USA*}

\author{Roman I. Vasquez}
\affiliation{Department of Physics, University of Central Florida, 4111 Libra Drive, Orlando, Florida, 32816-2385, USA}
\affiliation{Department of Mathematics and Statistics, Auburn University, 221 Roosevelt Concourse, Auburn, Alabama 36849, USA*}

\author{Andrew C. Silverman}
\affiliation{Department of Physics, University of Central Florida, 4111 Libra Drive, Orlando, Florida, 32816-2385, USA}

\author{Christopher L. Cain}
\affiliation{Department of Physics, University of Central Florida, 4111 Libra Drive, Orlando, Florida, 32816-2385, USA}
\affiliation{Engineering Technology, Mechatronics, St. Petersburg College, Clearwater, Florida 37765, USA*}

\author{Richard A. Klemm$^{\dag}$}
\affiliation{Department of Physics, University of Central Florida, 4111 Libra Drive, Orlando, Florida, 32816-2385, USA}
\affiliation{U. S. Research Laboratory, Wright-Patterson Air Force Base, Ohio, 45433-7251, USA}

\date{\today}

\begin{abstract}
For a spinless quantum  particle in a one-dimensional box or an electromagnetic wave in a one-dimensional cavity, the respective Dirichlet and Neumann boundary conditions both lead to non-degenerate wave functions.  However, in two spatial dimensions, the symmetry of the box or microstrip antenna is an important feature that has often been overlooked in the literature. In the high-symmetry cases of a disk, square, or equilateral triangle, the wave functions for each of those two  boundary conditions are grouped  into  two distinct classes, which are one- and two-dimensional representations of the respective point groups, $C_{\infty v}$, $C_{4v}$, and $C_{3v}$.  Here we present visualizations of representative wave functions for both boundary conditions and both one- and two-dimensional representations of those point groups.  For the one-dimensional representations, color contour plots of the wave functions are presented.  For the two-dimensional representations, the infinite degeneracies are presented as common nodal points and/or lines, the patterns of which are invariant under all operations of the respective point group.   The wave functions with the Neumann boundary conditions have important consequences for the coherent terahertz emission from the intrinsic Josephson junctions in the high-temperature superconductor Bi$_2$Sr$_2$CaCu$_2$O$_{8+\delta}$:  the enhancement of the output power from electromagnetic cavity resonances is only strong for wave functions that are not degenerate.
\end{abstract}

\maketitle 

\noindent{*present address}
\noindent{$^{\dag}$corresponding author.  email:  richard.klemm@ucf.edu}

\section{Introduction}
The study of wave functions obtained from various geometries with Dirichlet and Neumann boundary conditions has been a useful educational resource and also has numerous applications in the construction of various devices.  A quantum particle in a one-dimensional infinite square well potential, or ``box'', for which the boundary is a set of two points, is often the first problem studied by undergraduate students early in their first course on quantum mechanics \cite{GS}.  Although two graduate texts contain two problems on the degeneracies of the lowest energies of a square two-dimensional (2D) box \cite{Sakurai,SN}, the solutions manuals for those texts  both give an incorrect value for the  degeneracy of the square box's first excited state. Hence, there could be considerable confusion on this issue.  Here we focus upon high-symmetry, 2D  shapes.  For a spinless quantum particle of mass $M$ in a 2D infinite square well potential or box, the wave function $\psi(x,y,t)$ satisfies the Schr{\"o}dinger equation,
\begin{eqnarray}
-\frac{\hbar^2}{2M}{\bm\nabla}^2\psi+V\psi&=&i\hbar\frac{\partial\psi}{\partial t},
\end{eqnarray}
 where $\hbar=\frac{h}{2\pi}$ and $h$ is Planck's constant, for which the potential $V(x,y)=0$ inside the box and $V(x,y)=\infty$ outside it.  Hence $\psi(x,y,t)=0$ outside the box and on its boundary,
  the simplest example of Dirichlet boundary conditions.  Here we only consider closed one-dimensional boundaries, and focus upon the 2D shapes with the highest point-group symmetries, $C_{\infty v}$, $C_{4v}$, and $C_{3v}$, corresponding to  cylindrical, square, or equilateral triangular boxes \cite{Tinkham}.

   For a thin (nearly 2D) microstrip antenna (MSA),  the magnetic vector potential $A_z(x,y,t)$ normal to the antenna satisfies the electromagnetic (EM) wave equation
  \begin{eqnarray}
  {\bm\nabla}^2A_z-\frac{1}{v^2}\frac{\partial^2A_z}{\partial t^2}&=&0,
  \end{eqnarray}
  where $v$ is the wave velocity that depends upon the index of refraction in the antenna, and for transverse magnetic (TM) modes, its normal derivative vanishes on the boundary,
  the simplest example of Neumann boundary conditions.  Due to the oscillatory time dependence of a light wave, Eq. (2) is usually rewritten as ${\bm\nabla}^2A_z+(k')^2A_z=0$, where $k'$ is the wave vector in the material of interest.

In particular, determining the symmetries and energy states of wave functions with Neumann boundary conditions is of great practical importance in the development of a high-power terahertz (THz) laser, which has many potential applications, such as for the detection of skin or colon cancer in humans and in secure communications.  This is due to the ac Josephson effect, in which a dc voltage $V$ is applied across a single junction, leading to an ac current $I$ and the emission of photons at the frequency
$f_J=2eV/h$,
where $e$ is the charge of an electron \cite{Josephson}.  Now there exist many layered superconductors \cite{Klemmbook}, a number of which exhibit Josephson effects, but the most interesting one for the construction of a THz laser is   the high transition temperature $T_c$ superconductor Bi$_2$Sr$_2$CaCu$_2$O$_{8+\delta}$ (Bi2212).  This material consists of a uniform stack of  intrinsic Josephson junctions (IJJs) \cite{Kleiner1992,Kleiner1994,Batov2006,Ozyuzer2007}. The output power $P_1$ of a single IJJ is about 1 pW, too small for most practical  applications that require the actual power $P$ to be at least 1 mW.   But since each IJJ is 1.533 nm thick, a single crystal of Bi2212 of thickness 1 $\mu$m contains $N \approx 650$ IJJs, reducing  $f_J$  to
\begin{eqnarray}
f_{J}&=&(2e/h)(V/N),\label{Njunctions}
\end{eqnarray}
and when most of the $N$ junctions emit coherently, ideally, $P_N=P_1N^2\approx0.4 \mu$W. Moreover, in a thin mesa cut from a single crystal of Bi2212, the shape of the mesa acts as an electromagnetic cavity or MSA, which can enhance the output power an additional one to two orders of magnitude \cite{Ozyuzer2007,Kadowaki2010,Klemm2010a,Klemm2010b,Tsujimoto2012}.  The emission frequency has to be larger than the Josephson plasma resonance frequency $f_p\approx 0.25$ THz \cite{Singley2004}. In principle, the maximum emission frequency is the low temperature $T$ value of the superconducting gap $2\Delta$, which is about 15 THz \cite{Hoogenboom,Borodianskyi2017,Xue2016,Xue2020}.

 However, a major issue affecting the reliability and the upper limit of the emission frequency  has been the Joule heating of the mesas\cite{Wang2009,Wang2010,Minami2014}, especially when they have Bi2212 substrates, since Bi2212 is a very poor thermal conductor. But this problem has been mostly removed by fabricating stand-alone mesas, in which the Bi2212 sample is doubly cleaved to a thickness of 1-2 microns from a single crystal, and the top and bottom surfaces are each coated with a thin layer of gold \cite{Klemm2010a,Klemm2010b,Tsujimoto2012,Klemm2010c,Kitamura2014,Kashiwagi2015b,Kashiwagi2018,Kashiwagi2017b}.  These issues are discussed in Section VIII.  When different parts of a small stand-alone Bi2212 single crystal were used both as the emitter and as the detector,  emission up to 11 THz was observed \cite{Borodianskyi2017}.  An array of three stand-alone rectangular mesas was reported to emit coherently with the combined $P\approx 0.61$ mW \cite{Benseman2013}, although an array of two rectangular mesas did not behave coherently \cite{Kleiner2015}, so the main concern for further development is the design of arrays to increase that number.

For low-symmetry rectangular boxes or MSAs, the wave functions are all nondegenerate, one-dimensional representations (1DRs) of the point group $C_{2v}$.  However, for the higher symmetry  square \cite{Klemm2017,Klemm2017b,Sun2018,Ono2020},  equilateral triangular \cite{Helszajn1978,Overfelt1986,Delfanazari2012,Delfanazari2013,Delfanazari2013b,Klemm2013,Cerkoney2017}, isosceles and right triangular \cite{Delfanazari2012b,Delfanazari2014}, regular pentagonal \cite{Delfanazari2015}, cylindrical, disk  \cite{Tsujimoto2010,Kashiwagi2015}, annular \cite{Bonnough2018}, or singly slitted annular\cite{Shouk2021} shapes, the situation can be considerably more complicated. Although some of the wave functions are non-degenerate 1DRs of the respective point groups $C_{\infty v}$, $C_{4v}$ and $C_{3v}$ under consideration here, a large fraction of the wave functions are infinitely-degenerate two-dimensional representations (2DRs) of those point groups \cite{Tinkham}.  In addition, for square boxes or MSAs, many additional wave functions are doubly degenerate.   Previous works have calculated the dimensionality of the symmetry groups for the stationary states of the various wave functions, but there have not been thorough investigations of the features of the 2DRs of all three of these cases.  Although for 2D boxes, such considerations are only experimentally relevant for deep quantum wells, for nearly 2D MSAs, the experimental consequences are very important, but have not been clearly described in the literature.

We note that circularly polarized coherent THz emission can be obtained by breaking the symmetry of square or disk MSAs \cite{Asai2017,Elarabi2017,Elarabi2018}, Bi2212 MSAs can be used both as emitters and as detectors \cite{Borodianskyi2017,Irie2012}, and commercial cryocoolers can be used in cooling Bi2212 MSAs for many potential applications \cite{Nakade2014,Saiwai2020}.  To date, six review articles on Bi2212 IJJ-THz emitters have been published \cite{Kashiwagi2012,Welp2013,Kakeya2016,Kashiwagi2017,Kleiner2019,Delfanazari2020}.

 Here we present detailed studies of the wave functions for the three highest-symmetry 2D shapes:  cylindrical boxes and disk MSAs, square boxes and MSAs, and equilateral triangular boxes and MSAs, in which either the wave function or its normal derivative vanishes on the boundary.  Character tables of the respective point groups $C_{\infty v}$, $C_{4v}$ and $C_{3v}$ are given in textbooks on group theory \cite{Tinkham}, but there are some minor differences in the wave function tables with Dirichlet and Neumann boundary conditions we present here, and some additions for the 2DR wave functions.  In Section II, we analyze the  square box.  In Section III, we describe the thin square MSA. In Section IV, we analyze the equilateral triangular box.  In Section V, we present the results for thin equilateral triangular MSA wave functions.  In Section VI, we show the cylindrical box wave functions.  In Section VII, we describe the disk MSA wave functions. In Section VIII, we compare our results for  square, equilateral triangular, and disk MSAs with published experimental results.
 Finally in Section IX, we summarize our results for these high-symmetry boxes and MSAs.

\section{The square box}
For a spinless quantum particle of mass $M$ in a square box of side $a$, the normalized wave functions are solutions of the Schr{\"o}dinger equation with $V(x,y)=0$ for $0<x,y<a$, $V(x,y)=\infty$ for $x,y\le0$ and $x,y\ge a$, which are
\begin{eqnarray}
\Psi_{n,m}(x,y)&=&\frac{2}{a}\sin(n\pi x/a)\sin(m\pi y/a),\label{boxwavefunctions}
\end{eqnarray}
for integral $n,m\ge 1$, all of which satisfy the Dirichlet boundary conditions $\Psi_{n,m}(x,y)=0$ for $x=0,a$ and $y=0,a$.  The energy of that state is
\begin{eqnarray}
E_{n,m}&=&\frac{\hbar^2(n^2+m^2)\pi^2}{2Ma^2}.
\end{eqnarray}

Figure 1 displays color-coded plots of the wave functions  $\Psi_{n,m}(x,y)$ for  $1\le n,m\le 4$ of the square box.

\begin{figure}[h]
\center{\includegraphics[width=0.45\textwidth]{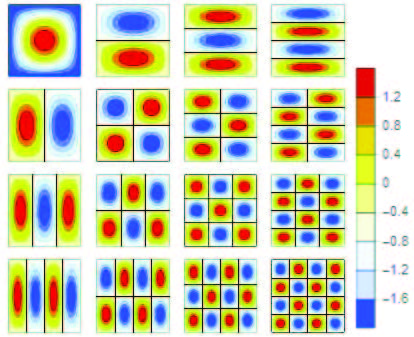}
\caption{(color online) Table of color-coded plots of the normalized wave functions $\Psi_{n,m}(x,y)=\frac{2}{a}\sin(n\pi x/a)\sin(m\pi y/a)$ for a square box of side $a$ with Dirichlet boundary conditions for $(n,m)=1,2,3,4$, each with its lower left corner at (0,0). Here we set $a=1$. $n$ and $m$ are respectively the column and row numbers. The color code bar applies to each of these figures. The black boundaries and internal lines are nodes. }}
\end{figure}

  According to the $C_{4v}$ point group symmetry class, there are four mirror planes:  the horizontal $\sigma_h$ and vertical $\sigma_v$ mirror planes that bisect the sides, and  the two diagonal mirror planes $\sigma_{d1}$ and $\sigma_{d2}$ that bisect the corners.  In addition,  two rotations ($R_4$) by $2\pi/4$ and one ($R_2$) by $\pi$ about the centroid are also allowed \cite{Tinkham}. The wave functions fall into three basic classes.  In the first class, $n=m$, $\Psi_{n,n}(x,y)$ is nondegenerate.  But there are two subclasses of these nondegenerate wave functions.  For $n$ odd, the $\Psi_{n,n}(x,y)$ are invariant under all of these operations, whereas for $n$ even, the $\Psi_{n,n}(x,y)$ are even about $\sigma_{d1},\sigma_{d2}$ and under $R_2$, but are odd about $\sigma_h,\sigma_v$ and under $R_4$.  According to Table I, the odd and even $n$ $\Psi_{n,n}(x,y)$ are respectively elements of subgroups $A_1$ and $B_2$.

\begin{figure}[h]
\center{\includegraphics[width=0.35\textwidth]{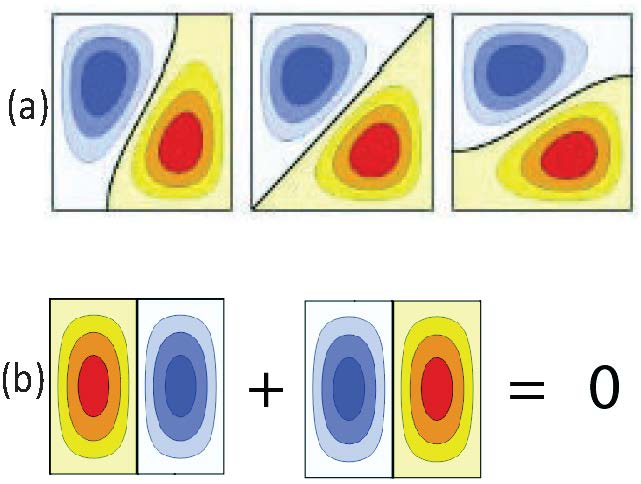}
\caption{(color online)(a) Color-coded plots of the infinitely-degenerate square box wave function $\Psi^{-\theta,+}_{1,2}(x,y)=\cos\theta\Psi_{1,2}(x,y)-\sin\theta\Psi_{2,1}(x,y)$ for $\phi=0$, $\theta=30^{\circ}$ (left), $45^{\circ}$ (center), and $60^{\circ}$ (right). (b) Color-coded illustration of the oddness of $\Psi_{2,1}(x,y)$ under rotations about its centroid at $(\frac{a}{2},\frac{a}{2})$  by $\pi$. }}
\end{figure}

  For $m=n+2p+1$, the pair of wave functions $\Psi_{n,n+2p+1}(x,y)$ and $\Psi_{n+2p+1,n}$ are odd under $R_2$, and have $E=2$, the trace of the rank 2 identity matrix ${\bm 1}$.  The full degeneracy may be represented by
  \begin{eqnarray}
  \Psi^{(\theta,+)}_{n,n+2p+1}&=&\cos\theta \Psi_{n,n+2p+1}(x,y)\nonumber\\&&+e^{i\phi}\sin\theta \Psi_{n+2p+1,n}(x,y),\nonumber\\
  \Psi^{(\theta,-)}_{n,n+2p+1}&=&-\sin\theta \Psi_{n,n+2p+1}(x,y)\nonumber\\&&+e^{i\phi}\cos\theta \Psi_{n+2p+1,n}(x,y),\label{2DRwavefunctions}
  \end{eqnarray}
  which are the two orthonormal subsets of the degenerate $\Psi_{n,n+2p+1}$ and $\Psi_{n+2p+1,n}$  wave functions \cite{Klemm2017b}.  However, since $0\le\theta<2\pi$, $\theta$ can be any real mixing angle, this degeneracy is infinite.  For simplicity, we shall assume that $\phi=0$, so that the wave functions are all real. This is entirely analogous to a spin $\frac{1}{2}$ system in the absence of a magnetic field \cite{SN}. In Fig. 2(a), this degeneracy is displayed for $\Psi_{1,2}^{-\theta,+}(x,y)$ for $\phi=0$ at the three mixing angles $\theta=30^{\circ}, 45^{\circ}$, and $60^{\circ}$.  In Fig. 2(b), the fact that such wave functions are odd under $R_2$ is evident for $\Psi_{2,1}(x,y)$ by adding $\Psi_{2,1}(x)+R_2\Psi_{2,1}(x,y)$, which  vanishes.
  The only points at which such wave functions are invariant under all of the operations of $C_{4v}$ are their common nodal points.  These wave functions are thus displayed for $1\le n,m\le 4$ as the appropriate sets of nodal points in Fig. 3.  We note that in each of these cases, the complete set of nodal points is invariant under each of the operations of $C_{4v}$ (all four mirror planes and both rotations).

  Writing these 2DR wave functions in the Nambu representation,
  \begin{eqnarray}
  \Psi^{\theta}_{n,n+2p+1}(x,y)&=&\Biggl(\begin{array}{c}\Psi^{(\theta,+)}_{n,n+2p+1}(x,y)\\ \Psi^{(\theta,-)}_{n,n+2p+1}(x,y)\end{array}\Biggr),\label{squarebox2DR}
  \end{eqnarray}
  where the $\Psi^{(\theta,\pm)}_{n,n+2p+1}(x,y)$ are given by Eq. (\ref{2DRwavefunctions}),
   the operations of $C_{4v}$ upon them are rank-2 matrices.  It is easy to show that $R_2=-{\bm 1}$, the  trace of which is -2.  The other operations can be written in terms of the Pauli matrices $\sigma_x$, $\sigma_y$, and $\sigma_z$. In particular, the 2 $R_4$ matrices are $\pm i\sigma_y$, $\sigma_h=\sigma_v=\sigma_z\cos(2\theta)-\sigma_x\sin(2\theta)$, and $\sigma_{d1}=-\sigma_{d_2}=\sigma_z\sin(2\theta)+\sigma_x\cos(2\theta)$.  Since the Pauli matrices are traceless, the traces of the 2 $R_4$, $\sigma_v$, $\sigma_h$, $\sigma_{d1}$ and $\sigma_{d2}$ all vanish for any value of $\theta$.

 \begin{figure}
\center{\includegraphics[width=0.42\textwidth]{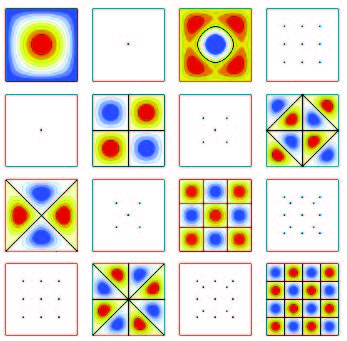}
\caption{(color online) Table of color-coded plots of the square box non-degenerate $\Psi_{n,n}(x,y)$  along the table diagonal, the doubly-degenerate $\Psi^{+}_{n,n+2p}(x,y)$ and $\Psi^{-}_{n,n+2p}(x,y)$ are displayed in the regions above and below the diagonal, respectively, and the infinitely degenerate $\Psi_{n,n+2p+1}^{\theta,\pm}(x,y)$ are represented by their common nodal points.  Each such set of common nodal points is a one-dimensional representation of $C_{4v}$ that is even under each of its symmetry operations:  $\sigma_x,\sigma_y,\sigma_{d1},\sigma_{d2},R_4$, and $R_2$. The constant contours of the 1DR wave functions nearest to  the boundary are parallel to it, due to the Dirichlet boundary conditions. }}
\end{figure}

\begin{table}
\vskip10pt
\begin{tabular}{lccrrrrrr}
\hline
\noalign{\vskip2pt}
& &&&&&&$\sigma_h$,&$\sigma_{d1}$,\\
Type&Symmetry&$\psi^{(\pm)}_{n,m}(x,y)$&$n$&$E$&$R_2$&$2R_4$&$\sigma_v$&$\sigma_{d2}$\\
\noalign{\vskip3pt}
\hline
\noalign{\vskip3pt}
$A_1$&$x^2+y^2$&$\Psi_{n,n},\Psi_{n,n+2p}^{(+)}$&o&1&+1&+1&+1&+1\\
$A_2$&$xy(x^2-y^2)$&$\Psi_{n,n+2p}^{(-)}$&e&1&+1&+1&-1&-1\\
$B_1$&$x^2-y^2$&$\Psi^{(-)}_{n,n+2p}$&o&1&+1&-1&+1&-1\\
$B_2$&$xy$&$\Psi_{n,n},\Psi_{n,n+2p}^{(+)}$&e&1&+1&-1&-1&+1\\
$A_1$*&point nodes&$\Psi^{(\theta,\pm)}_{n,n+2p+1}$&e,o&2&-2&0&0&0\\
$A_1$*&nodal squares$^{\dag}$&$\Psi^{(\theta,\pm)}_{n,n+2p+1}$&e,o&2&-2&0&0&0\\
\noalign{\vskip3pt}
\hline
\noalign{\vskip3pt}
\end{tabular}
\caption{Square box representation types, symmetries, allowed 1DRs  $\Psi_{n,n}(x,y)$ and $\Psi_{n,n+2p}^{(\pm)}(x,y)$ [Eq. (7)] for odd or even $n\ge1$, 2DRs $\Psi^{(\theta,\pm)}_{n,n+2p+1}(x,y)$ [Eq. (6)] of the square box, and operations of the $C_{4v}$ point group. $\sigma_h$, $\sigma_v$, $\sigma_{d1}$, and $\sigma_{d2}$ are the mirror planes along the horizonal and vertical axes and the two diagonals,  $R_n$ represents rotations by $2\pi/n$ about the centroid, and $E$ is the trace of the identity matrix for the appropriate group dimension. For the 2DR wave functions, the listed elements are the traces of the rank-2 matrices that describe the operations \cite{Tinkham}. *Note that for the 2DR wave functions with no symmetry, their sets of common point nodes and square nodes all have $A_1$ symmetry.  $^{\dag}$Under special conditions.  See Figs. 3 and 4 and the text.}
\end{table}

\begin{figure}
\center{\includegraphics[width=0.45\textwidth]{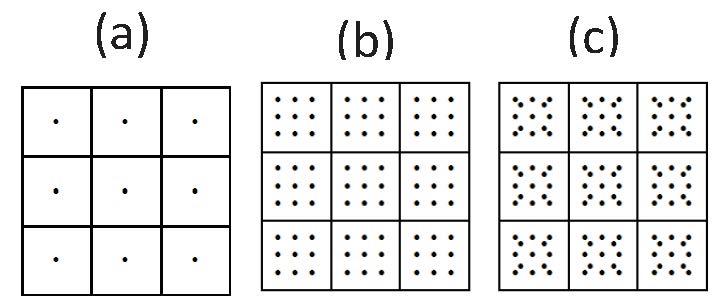}
\caption{Presentations of selected 2DRs with dots and squares. (a) $\Psi_{3,6}(x,y),\Psi_{6,3}(x,y)$. (b) $\Psi_{3,12}(x,y),\Psi_{12,3}(x,y)$. (c) $\Psi_{9,12}(x,y),\Psi_{12,9}(x,y)$.}}
\end{figure}

 For $m=n+2p$, there are again two classes.  For $n$ odd, $\Psi_{n,n+2p}(x,y)$ is even under reflections about $\sigma_h$, $\sigma_v$ and under $R_2$, but shows no symmetry under $R_4$ and about $\sigma_{d1}$, $\sigma_{d2}$.    For $n$ even, $\Psi_{n,n+2p}(x,y)$ is odd under reflections about $\sigma_x$, $\sigma_y$ and under $R_2$, but shows no symmetry under $R_4$ and about $\sigma_{d1}$, $\sigma_{d2}$.  However, we note that for both $n$ odd or even, $R_4\Psi_{n,n+2p}(x,y)=\Psi_{n+2p,n}(x,y)$.  This implies that there are two orthonormal members of each subgroup,
 \begin{eqnarray}
\Psi^{\pm}_{n,n+2p}(x,y)&=&[\Psi_{n,n+2p}(x,y)\pm\Psi_{n+2p,n}(x,y)]/\sqrt{2},\label{doublydegenerate}
\end{eqnarray}
which are doubly degenerate.
These wave functions are also displayed in Fig. 3.

We note that $\Psi_{n,n+2p}^{+}(x,y)$ for $n$ odd is invariant under all $C_{4v}$ operations, as seen in Fig. 3 for $\Psi_{1,3}^{+}(x,y)$.  Hence, such wave functions are in symmetry group $A_1$, and are listed as such in Table I.  For $n$ even, $\Psi^{+}_{n,n+2p}(x,y)$ is even under reflections about $\sigma_{d1}$, $\sigma_{d2}$ and under $R_2$, but is odd under reflections about $\sigma_x$, $\sigma_y$ and under the 2 $R_4$, so it is in symmetry group $B_2$.  For $n$ odd, $\Psi^{-}_{n,n+2p}(x,y)$ is even under reflections about $\sigma_h$, $\sigma_v$ and $R_2$, but odd under reflections about $\sigma_{d1}$, $\sigma_{d2}$ and under $R_4$, so it is in symmetry group $B_1$. Finally, for $n$ even, $\Psi^{-}_{n,n+2p}(x,y)$ is odd under reflections about all four mirror planes and under $R_4$, but is even under $R_2$, so it is in symmetry group $A_2$.  These classifications are all listed in Table I.

In addition to the common nodal point structure of each infinitely degenerate 2DR wave function, some higher index 2DR wave functions have mutual nodal squares, some examples of which are shown in Fig. 4.  For such common nodal squares to appear in the common nodal set of 2DRs, the lower quantum number  $n\ge3$ must be odd.  The lowest energy case is therefore the (3,6) case pictured in Fig. 4(a).  In Fig. 4(b), the (3,12) case is shown.  In addition, a more complicated nodal pattern is obtained for the (9,12) case pictured in Fig. 4(c).  In this case, both numbers factor into 3 times an odd or an even number, and this factorization allows for the square common nodal structure, each of which encloses a finite set of common nodal dots.  Obviously as the lower odd number increases, the common nodal patterns  become increasingly complicated.  But it is noteworthy that in every set of common nodal dots and/or squares, that set is invariant under all operations of $C_{4v}$, and hence obeys the  $A_1$ symmetry table.

\section{The square microstrip antenna}
For a thin square microstrip antenna of the same geometry as for the square box, with its origin at the lower left corner, but satisfying the EM wave equation for $A_z(x,y,t)$, the normalized transverse magnetic wave functions at a fixed time with the appropriate Neumann boundary conditions,
\begin{eqnarray}
\frac{\partial\Psi_{n,m}(x,y)}{\partial x}\Bigl|_{x=0,a}&=&\frac{\partial\Psi_{n,m}(x,y)}{\partial y}\Bigl|_{y=0,a}=0,\nonumber\\
\end{eqnarray}
 have the form
\begin{eqnarray}
\Psi_{n,m}(x,y)&=&\left\{\begin{array}{lc}\frac{2}{a}\cos(n\pi x/a)\cos(m\pi y/a),&m,n\ge1\\
\frac{\sqrt{2}}{a}\cos(n\pi x/a),&m=0,n\ge1\\
\frac{\sqrt{2}}{a}\cos(m\pi y/a),&n=0,m\ge1\end{array}\right.\nonumber\\
&&\label{MSAwavefunctions}
\end{eqnarray}
 The enhanced emission frequencies $f_{n,m}$ from the square thin MSA are
\begin{eqnarray}
f_{n,m}&=&c_0\frac{\sqrt{n^2+m^2}}{2an_{\rm r}},
\end{eqnarray}
where $c_0$ is the vacuum speed of light and $n_{\rm r}$ is the index of refraction, which for Bi2212 IJJ-THz emitter devices that are on the order of 1 $\mu$m thick, is about 4.2.  The case $n=m=0$ must be excluded, since light must have a finite frequency.
In this case, the color-coded lowest frequency square MSA wave functions are shown in Fig. 5.
\begin{figure}
\center{\includegraphics[width=0.48\textwidth]{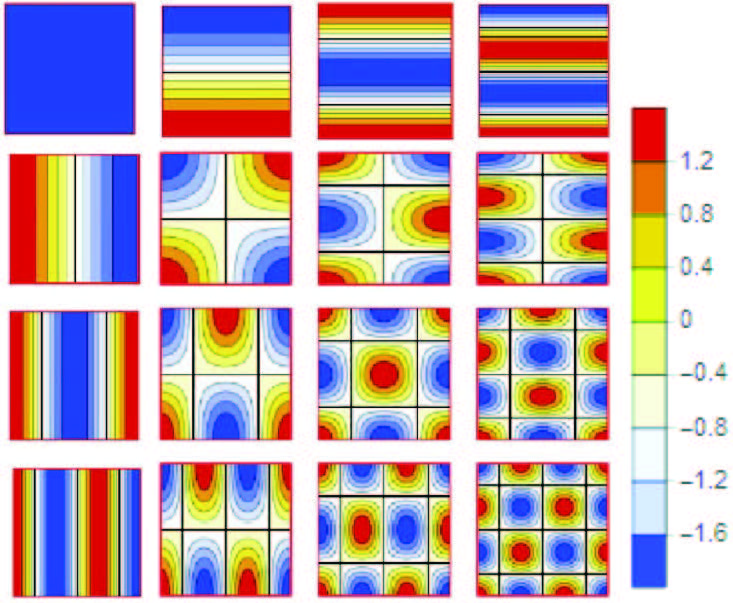}
\caption{(color online) Color-coded plots of the square MSA $\Psi_{n,m}(x,y)$ given by Eq. (\ref{MSAwavefunctions}) for $0\le n,m\le 3$, which are the nominal representations of the lowest frequency wave functions of a square microstrip antenna with Neumann boundary conditions that are indicated by the red boundaries.  $n$ and $m$ are respectively the column and row numbers. $\Psi_{0,0}=1/a$  shown in blue corresponds to frequency  $f_{0,0}=0$, so it is excluded. The color code bar otherwise applies to each of these figures.}}
\end{figure}
\begin{table}
\vskip10pt
\begin{tabular}{lccrrrrrr}
\hline
\noalign{\vskip2pt}
& &&&&&&$\sigma_h$,&$\sigma_{d1}$,\\
Type&Symmetry&$\psi^{(\pm)}_{n,n'}(x,y)$&$n$&$E$&$R_2$&$2R_4$&$\sigma_v$&$\sigma_{d2}$\\
\noalign{\vskip2pt}
\hline
\noalign{\vskip3pt}
$A_1$&$x^2+y^2$&$\Psi_{n,n},\Psi_{n,n+2p}^{(+)}$&e&1&+1&+1&+1&+1\\
$A_2$&$xy(x^2-y^2)$&$\Psi_{n,n+2p}^{(-)}$&o&1&+1&+1&-1&-1\\
$B_1$&$x^2-y^2$&$\Psi^{(-)}_{n,n+2p}$&e&1&+1&-1&+1&-1\\
$B_2$&$xy$&$\Psi_{n,n},\Psi_{n,n+2p}^{(+)}$&o&1&+1&-1&-1&+1\\
$A_1$&point nodes*&$\Psi^{(\theta,\pm)}_{n,n+2p+1}$&e,o&2&-2&0&0&0\\
\noalign{\vskip3pt}
\hline
\noalign{\vskip3pt}
\end{tabular}
\caption{Square MSA representation types, symmetries, allowed 1DRs  $\Psi_{n,n}(x,y)$ and $\Psi_{n,n+2p}^{(\pm)}(x,y)=[\Psi_{n,n+2p}(x,y)\pm\Psi_{n+2p,n}(x,y)]/\sqrt{2}$ for odd or even $n\ge1$, 2DRs $\Psi^{(\theta,\pm)}_{n,n+2p+1}(x,y)$, which have the forms of Eq. (6), except that their components satisfy Eq. (9), and operations of the $C_{4v}$ point group. For the 2DR wave functions, the listed values are the traces of the rank-2 matrices that describe the operations \cite{Tinkham}. *Note that the 2DR wave functions have no symmetry, but their sets of common fixed point nodes have $A_1$ symmetry. See text \cite{Klemm2017b}.}
\end{table}

As for the square box wave functions, the $n=m$ MSA wave functions are all non-degenerate 1DRs of the $C_{4v}$ point group, and the $m=n+2p+1$ MSA  wave functions are infinitely degenerate 2DRs of $C_{4v}$, having the forms of Eqs. (\ref{2DRwavefunctions}) and (\ref{squarebox2DR}), except that the wave function components are given by Eq. (\ref{MSAwavefunctions}) instead of Eq. (\ref{boxwavefunctions}).  This second point is illustrated in Fig. 6. In addition, since the MSA wave functions satisfy $\Psi_{n,n+2p}(x,y)=R_4\Psi_{n+2p,n}(x,y)$, exactly as for the square box, these wave functions are likewise doubly degenerate, and satisfy Eq. (\ref{doublydegenerate}), although again with the wave function components given by
Eq. (\ref{MSAwavefunctions}). The symmetry table of the square MSA wave functions is therefore shown in Table II.  The only differences between Table II and Table I for the square box is that the oddness or evenness of the quantum number $n$ is precisely the opposite, and there are no nodal squares for the square MSA 2DR wave functions.  The matrices that describe the symmetry operations upon the 2DR MSA wave functions are identical to those presented for the square quantum box wave functions following Eq. (\ref{squarebox2DR}).

We then redisplay those and additional MSA wave functions in the array shown in Fig. 7.  As for the square box, the diagonal $n=m$ square MSA wave functions are all nondegenerate 1DRs of $C_{4v}$, and are displayed in color-coded contour plots.  The infinitely degenerate $m=n+2p+1$ cases are again displayed as a set of mutually common nodal points, but in this case, as discussed in more detail in the following, there are no mutual nodal lines at any higher index number.  Furthermore, the $m=n+2p$ cases are also doubly degenerate, with the $\Psi^{\pm}_{n,n+2p}(x,y)$ given by Eq. (\ref{doublydegenerate}) with the appropriate MSA wave functions, and with the $\Psi^{+}_{n,n+2p}$ displayed above the diagonal, and the $\Psi^{-}_{n,n+2p}$ displayed below the array diagonal, exactly as in Fig. 3 for the box.

\begin{figure}[h]
\center{\includegraphics[width=0.45\textwidth]{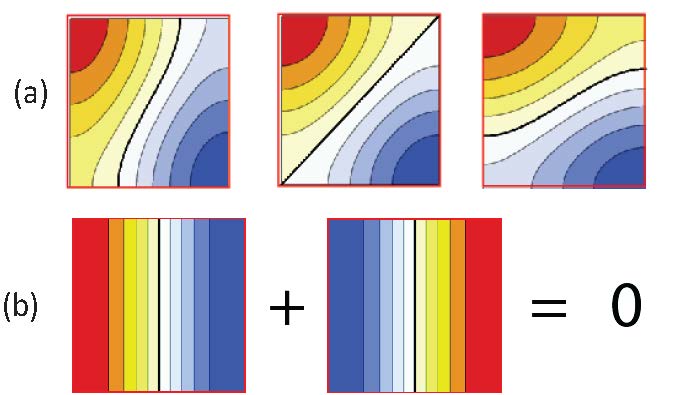}
\caption{(color online) (a) Color-coded plots of the square MSA 2DR $\Psi^{-\theta,+}_{0,1}(x,y)$ for $\theta=30^{\circ}, 45^{\circ}$, and $60^{\circ}$ from left to right. (b) Plot of $\Psi_{0,1}(x,y)+R_2\Psi_{0,1}(x,y)$, which equals 0.}}
\end{figure}

\begin{figure}[h]
\center{\includegraphics[width=0.45\textwidth]{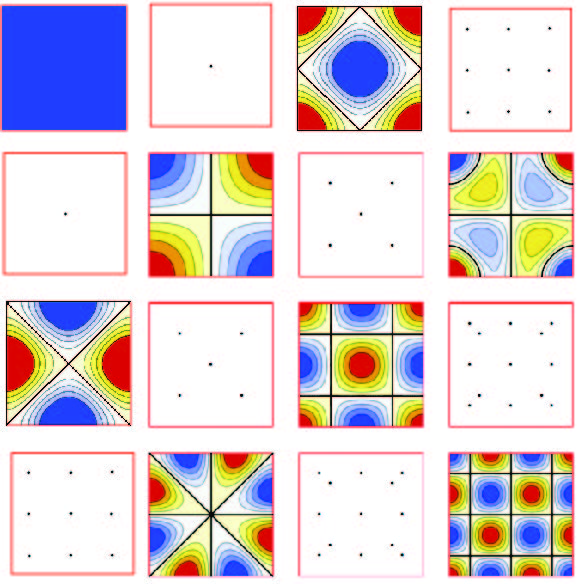}
\caption{(color online) Shown are plots of the array of square MSA wave functions $\Psi_{n,m}(x,y)$ with Neumann boundary conditions for $0\le n,m\le 4$.  1DR wave functions of nondegenerate $\Psi_{n,n}$ (for $n\ge1$) and doubly-degenerate $\Psi_{n,n+2p}^{\pm}$ are shown in color-coded plots along the array diagonal and in the appropriate positions above (below) that diagonal.  2DR Plots of the infinitely-degenerate $\Psi_{n,n+2p+1}$ are shown as arrays of black mutual nodal points. The constant contours nearest to a boundary intersect it normally due to the Neumann boundary conditions. }}
\end{figure}

 Fr square box 2DR wave functions with $n,m>0$  without any line nodes, the number of  point nodes is $N(n,m) = (n-1)^2+(m-1)^2$.
Similarly, for $n,m\ge0$, the number of square MSA  point nodes in a 2DR is $N(n,m) = n^2+m^2$.
However, as argued in the following, no nodal lines or squares appear in the square MSA 2DRs.  But as for the square box, the set of common nodal points for each 2DR is invariant under all operations of $C_{4v}$, and hence obeys the $A_1$ symmetry table.

A nodal line could occur along the $x$ direction if $\cos(n\pi x/a)=0$ for $0<x<a$ and along the $y$ direction if $\cos(m\pi y/a)=0$ for $0<y<a$.  These require $x/a=(2p+1)/(2n)$ and $y/a=(2q+1)/(2m)$ for integer $p$ and $q$.  For them to occur together and the pattern to be invariant under the operations of $C_{4v}$, including $R_4$, which interchanges $x$ and $y$, we then require
\begin{eqnarray}
\frac{2p+1}{2n}&=&\frac{2q+1}{2m}.\label{node}
\end{eqnarray}

As noted from Table II, if $n = m$,  $\Psi_{n,m}(x,y)$  is a non-degenerate 1DR, and if $n$ and $m$ are either both odd or both even, then
$\Psi_{n,m}(x,y)$ is a doubly-degenerate combination of two  1DRs.  But if either $n$ or $m$ is odd and the other is even, then $\Psi_{n,m}(x,y) $ is a 2DR.  Rewriting Eq. (11) as $(2p+1)m=(2q+1)n$, it is easy to see this cannot be satisfied with either $n$ or $m$ odd and the other even.

\section{The equilateral triangular box}

Previous studies have focused on the 1DR wave functions of equilateral triangular MSAs \cite{Cerkoney2017}. Here we calculate the quantum wave functions and normalization constants for the equilateral triangle of side $a$ in an infinite potential well. The Schr{\"o}dinger equation admits even and odd wave function solutions about any of the symmetry axes.  Here we choose the horizontal axis as the axis of symmetry from which the wave functions can be generated.  The equilateral triangular box wave functions even and odd about that axis may be written as

\begin{eqnarray}
\Psi^{e}_{\ell,m,n}(x,y)\!\!&=&\!\! A^e_{m,n}\biggl\{\sin\left[\left(\frac{2\pi x}{\sqrt{3}a}+\frac{2\pi}{3}\right)\ell\right]\nonumber\\
& &\times\cos\left[\frac{2\pi (m-n)y}{3a}\right] \nonumber\\
&& + \sin\left[\left(\frac{2\pi x}{\sqrt{3}a} + \frac{2\pi}{3}\right)m\right]\cos\left[\frac{2\pi (n-\ell)y}{3a}\right]\nonumber\\
&&+ \sin\left[\left(\frac{2\pi x}{\sqrt{3}a} + \frac{2\pi}{3}\right)n\right]\cos\left[\frac{2\pi (\ell-m)y}{3a}\right]\biggr\}\>\>\>\>\>\>\>\>\label{etboxeven}
\end{eqnarray}
and
\begin{eqnarray}
\Psi^{o}_{\ell,m,n}(x,y)\!\!& =&\!\!A^o_{m,n}\biggl\{ \sin\left[\left(\frac{2\pi x}{\sqrt{3}a}+\frac{2\pi}{3}\right)\ell\right]\nonumber\\
& &\times\sin\left[\frac{2\pi (m-n)y}{3a}\right] \nonumber \\
&& + \sin\left[\left(\frac{2\pi x}{\sqrt{3}a} + \frac{2\pi}{3}\right)m\right]\sin\left[\frac{2\pi (n-\ell)y}{3a}\right]\nonumber\\
&&+ \sin\left[\left(\frac{2\pi x}{\sqrt{3}a} + \frac{2\pi}{3}\right)n\right]\sin\left[\frac{2\pi (\ell-m)y}{3a}\right]\biggr\}.\>\>\>\>\>\>\>\>\label{etboxodd}
\end{eqnarray}
Since each of the three terms for the even and odd wave functions must satisfy the Schr{\"o}dinger equation $-\frac{\hbar^2}{2M}{\bm\nabla}^2\Psi=E\Psi=\frac{\hbar^2k^2}{2M}\Psi=0$, this is equivalent to the EM wave equation, ${\bm\nabla}^2\Psi+k^2\Psi=0$. Since for each wave function form, each term must separately satisfy that wave equation, we find
\begin{eqnarray}
(n-m)(\ell+n+m)&=&0,
\end{eqnarray}
as for the equilateral MSA \cite{Cerkoney2017}.  As for the antenna, the $n=m$ cases can be shown to not produce any additional wave functions, so we assume $\ell=-n-m$.  However, in this case, both the odd and even wave functions vanish on the entire equilateral triangular boundary.

The energies for a quantum particle of mass $M$ in an equilateral triangular quantum box are then found to be
\begin{eqnarray}
E_{n,m}&=&\frac{(4\pi\hbar)^2}{2M(3a)^2}(m^2+n^2+mn),
\end{eqnarray}
which is $\propto k^2$ for the equilateral triangular MSA.
The
corresponding normalization coefficients are obtained by integrating $|\Psi^{o,e}_{\ell,m,n}|^2$ with $\ell=-n-m$ over the area of the equilateral triangle, and dividing by that area.  We find
\begin{eqnarray}
A_{m,n}^{e}& = &\left\{
        \begin{array}{ll}
            \frac{4}{3^{3/4}a}, & \quad m>n\neq0, n>m\neq0 \\
            \frac{2\sqrt{2}}{3^{3/4}a}, & \quad m=n\neq0\hskip10pt,
        \end{array}
    \right.\nonumber\\
A_{m,n}^{o}& = &\frac{4}{3^{3/4}a},  \quad  m>n\neq0, n>m\neq0.\label{boxconstants}
\end{eqnarray}
These equilateral triangular box normalization constants are remarkably similar to those obtained for the equilateral triangular MSA \cite{Cerkoney2017}, as shown in Section IV.

\begin{figure}[h]
\center{\includegraphics[width=0.48\textwidth]{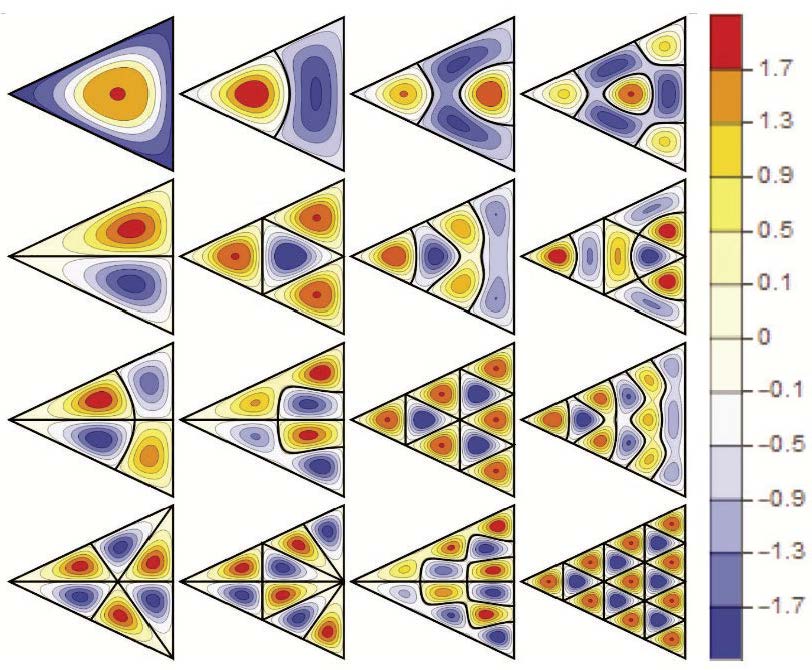}
\caption{(color online) Array of color coded plots of the lowest energy, normalized equilateral triangular box wave functions with $n,m=1,2,3,4$ from Eqs. (\ref{etboxeven}), (\ref{etboxodd}), and (\ref{boxconstants}).  The array diagonal consists of 1DR wave functions that are invariant under all $C_{3v}$ operations. The figures above and below the diagonal are respectively even and odd about the horizontal axis. The color code bar applies to each of these figures.}}
\end{figure}

\begin{table}
\vskip10pt
\begin{tabular}{lccccc}
\hline
\noalign{\vskip2pt}
Type&Symmetry&$\psi^{(e,o)}_{n,m}(x,y)$&$E$&$2R_3$&$3\sigma_v$\\
\noalign{\vskip3pt}
\hline
\noalign{\vskip3pt}
$A_1$&$x^2+y^2$&$\Psi^{e}_{n,n+3p}(x,y)$&1&+1&+1\\
$A_2$&$y(3x^2-y^2)$&$\Psi^{o}_{n,n+3p}(x,y)$&1&+1&-1\\
$A_1*$&nodal points&$\Psi^{(o,e)}_{n,m\ne n+3p}$,&2&-1&0\\
&&$n,m$ not both even&&&\\
$A_1*$&nodal points&$\Psi^{(o,e)}_{n,m\ne n+3p}$, $n,m$ both even&2&-1&0\\
& and triangles&&&&\\
\noalign{\vskip3pt}
\hline
\noalign{\vskip3pt}
\end{tabular}
\caption{Representation types, symmetries, allowed 1DRs  $\Psi^{e,o}_{n,n+3p}(x,y)$ and 2DRs $\Psi^{(o,e)}_{n,m\ne n+3p}(x,y)$  of the equilateral triangular box, and operations of the $C_{3v}$ point group. For 1DR wave functions, there  are 3 mirror planes $\sigma_v$ that bisect each angle, two rotations  $R_3$  by $\pm 2\pi/3$ about the centroid, and $E$ is the trace of the identity matrix for the appropriate group dimension \cite{Tinkham}. The 2DR wave functions have only one $\sigma_v$. *common nodal structure.  See Figs. (9) - (11) and the text.}
\end{table}
To illustrate  examples of the equilateral triangular box wave functions that are even or odd about only one vertex, we show pictorially that for 2DR equilateral triangular box wave functions that
\begin{eqnarray}
\!\!|\Psi^{(e,o)}_{2,3}(x,y)\rangle\!+\!R_3|\Psi_{2,3}^{(e,o)}(x,y)\rangle\!+\!R_3^2|\Psi^{(e,o)}_{2,3}(x,y)\rangle &\!=\!&0,\>\>\>\>\label{2DREpsis}
\end{eqnarray}
where we have used the Dirac ket notation.
\begin{figure}[h]
\center{\includegraphics[width=0.45\textwidth]{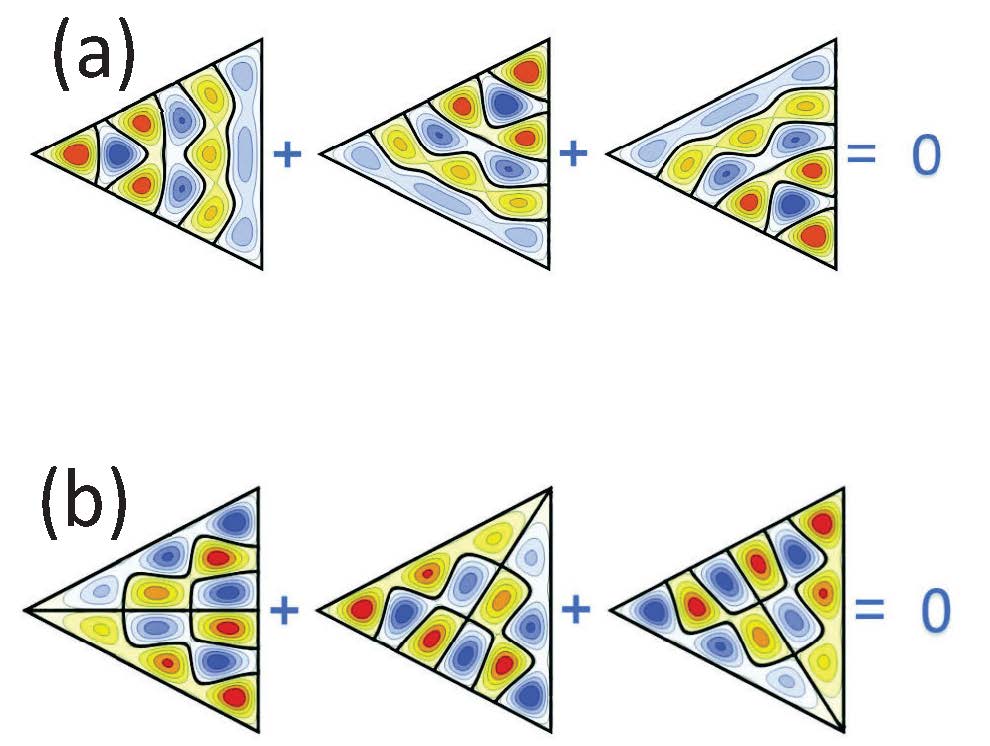}
\caption{(color online) Even (a) and odd (b) examples of 2DR equilateral triangular box wave functions pictured in Fig. 8 each satisfy the picture equation $|\Psi\rangle +R_3|\Psi\rangle+R_3^2|\Psi\rangle=0$.  In these examples, $(n,m)=(2,3)$.}}
\end{figure}

These equations show that only two of the wave functions even or odd about only one vertex are linearly independent, demonstrating that  these wave functions are 2DR wave functions.  Thus, we could choose as the general basis $|\Psi^{(e,o)}_{n,m}(x,y)\rangle$ and $R_3|\Psi^{(e,o)}_{n,m}(x,y)\rangle$, where $m\ne n+3p$.

 But since these wave functions are not orthonormal, we set
\begin{eqnarray}
|\Psi_{n,m}^{(e,o,1)}(x,y)\rangle&\!=\!&A|\Psi^{(e,o)}_{n,m}(x,y)\rangle\!+\!BR_3|\Psi^{(e,o)}(x,y)\rangle,\>\>\>\nonumber\\
|\Psi_{n,m}^{(e,o,2)}(x,y)\rangle&\!=\!&C|\Psi^{(e,o)}_{n,m}(x,y)\rangle\!+\!DR_3|\Psi^{(e,o)}(x,y)\rangle,\>\>\>
\end{eqnarray}
for constants $A, B, C$, and $D$,
and force them to form an orthonormal set.

To do so, we first take the inner product of Eq. (\ref{2DREpsis}) with
$\langle\Psi^{(e,o)}_{n,m}(x,y)|R_3^{\dag}$, and it is easily seen that
\begin{eqnarray}
\langle\Psi^{(e,o)}_{n,m}(x,y)|R_3^{\dag}\Psi^{(e,o)}_{n,m}(x,y)\rangle&=&-\frac{1}{2}.
\end{eqnarray}

Although complex coefficients are possible, especially for $|A|>\frac{2}{\sqrt{3}}$, under the assumption that all coefficients are real, it is then easy to show that the orthonormal set may be written as
\begin{eqnarray}
|\Psi_{n,m\ne n+3p}^{(e,o,1,\theta)}(x,y)\rangle&=&\frac{2}{\sqrt{3}}\cos\theta|\Psi^{(e,o)}_{n,m}(x,y)\rangle\nonumber\\& &+\Bigl(\frac{1}{\sqrt{3}}\cos\theta\pm\sin\theta\Bigr)R_3|\Psi^{(e,o)}_{n,m}(x,y)\rangle,\nonumber\\
|\Psi_{n,m\ne n+3p}^{(e,o,2,\theta)}(x,y)\rangle&\!=\!&-\frac{2}{\sqrt{3}}\sin\theta|\Psi^{(e,o)}_{n,m}(x,y)\rangle\nonumber\\&&\!-\Bigl(\frac{1}{\sqrt{3}}\sin\theta\mp\cos\theta\Bigr)R_3|\Psi^{(e,o)}_{n,m}(x,y)\rangle,\nonumber\\
&&\label{Phi2DREBox}\end{eqnarray}
where $0\le\theta<2\pi$ is arbitrary.  As for the square quantum box with $m=n+2p+1$ , the equilateral triangular quantum box wave functions with
$m\ne n+3p$ are 2DRs and are also infinitely degenerate.

When acting on the Nambu form of these 2DR wave functions,
\begin{eqnarray}
|\Psi^{(e,o,\theta)}_{n,m\ne n+3p}(x,y)\rangle&=&\Biggl(\begin{array}{c}|\Psi^{(e,o,1,\theta)}_{n,m\ne n+3p}(x,y)\rangle\\ |\Psi^{(e,o,2,\theta)}_{n,m\ne n+3p}(x,y)\rangle\end{array}\Biggr),\label{boxE2DR}
\end{eqnarray}
 the matrices $R_3$ and $R_3^{\dag}$ are

\begin{eqnarray}
R_3&=&-\frac{1}{2}{\bm 1}\pm i\frac{\sqrt{3}}{2}\sigma_y,
\end{eqnarray}
and its Hermitian conjugate,  both  traces of which are -1, as indicated in Table III.

With regard to the mirror symmetry operations of a 2DR wave function about a single vertex, the even  wave functions satisfy $\sigma^{(e)}_{v}|\Psi^{(e,\theta)}_{n,m}\rangle=|\Psi^{(e,\theta)}_{n,m}\rangle$, $\sigma^{(e)}_{v}R_3|\Psi^{(e,\theta)}_{n,m}\rangle=R_3^2|\Psi^{(e,\theta)}_{n,m}\rangle$ and $\sigma^{(e)}_{v}R_3^2|\Psi^{(e,\theta)}_{n,m}\rangle=R_3|\Psi^{(e,\theta)}_{n,m}\rangle$, as evidenced from Fig. 9(a).  Combining these equations with Eq. (\ref{2DREpsis}),  it is then straightforward to show that the mirror plane matrix $\sigma_v^{(e)}$ when acting on the Nambu form of Eq. (\ref{Phi2DREBox}) may be written for the even functions as
\begin{eqnarray}
\sigma_{v}^{(e)}&=&\frac{\sigma_z}{2}\bigl[\cos(2\theta)\mp\sqrt{3}\sin(2\theta)\bigr]\nonumber\\&&-\frac{\sigma_x}{2}\bigl[\sin(2\theta)\pm\sqrt{3}\cos(2\theta)\bigr],\label{sigmaveven}
\end{eqnarray}
which is traceless, as indicated in Table III.  On other hand, the 2DR wave functions odd about one vertex satisfy $\sigma^{(o)}_{v}|\Psi^{(o,\theta)}_{n,m}\rangle=-|\Psi^{(o,\theta)}_{n,m}\rangle$, $\sigma^{(o)}_{v}R_3|\Psi^{(o,\theta)}_{n,m}\rangle=-R_3^2|\Psi^{(o,\theta)}_{n,m}\rangle$ and $\sigma^{(o)}_{v}R_3^2|\Psi^{(o,\theta)}_{n,m}\rangle=-R_3|\Psi^{(o,\theta)}_{n,m}\rangle$, as sketched in Fig. 9(b).  Again, combining these equations with Eq. (\ref{2DREpsis}), when acting upon the odd Nambu form of Eq. (\ref{Phi2DREBox}), $\sigma_v^{(o)}=-\sigma_v^{(e)}$, which is given by Eq. (\ref{sigmaveven}), so that both traces of $\sigma_v^{(e)}$ and $\sigma_v^{(o)}$ vanish, as indicated in Table III.

As for the square box, we then  redraw the 2DR equilateral triangular box wave functions in terms of their common set of nodes.  This results in the array pictured in Fig. 10.  We note that $\Psi_{1,4}^{e}(x,y)$ and $\Psi_{1,4}^{o}(x,y)$ pictured in the top right and bottom left array positions are 1DRs, as are all four of the $\Psi_{n,n}^{e}(x,y)$ along the array diagonal.  In addition, $\Psi_{2,4}^{e}(x,y)$ and $\Psi_{2,4}^{o}(x,y)$ are both 2DRs that contain an identical set of nodal points plus a single equilateral triangular nodal figure in their center.
\begin{figure}[h]
\center{\includegraphics[width=0.45\textwidth]{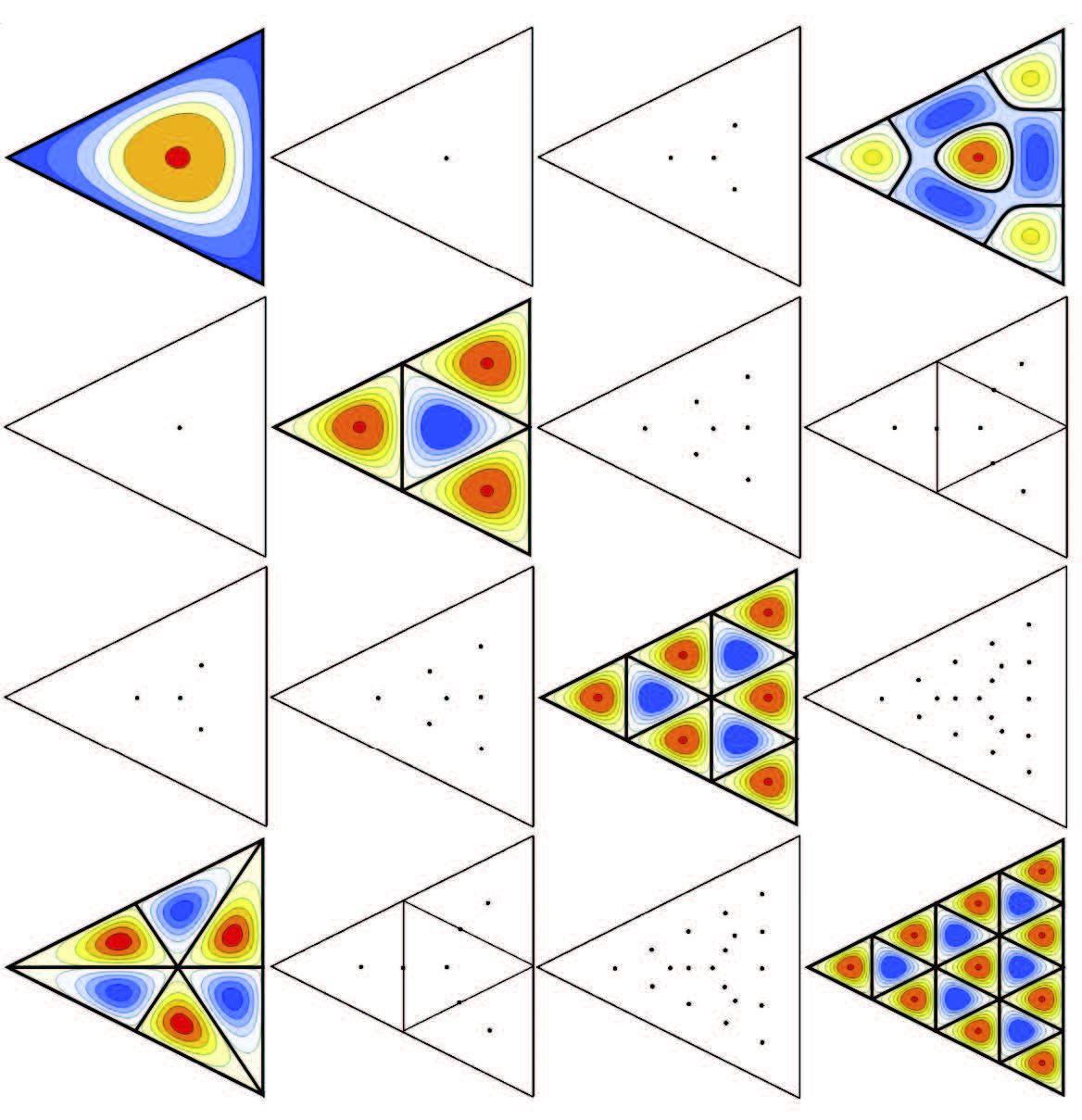}
\caption{(color online) Plots of the same equilateral triangular box wave functions pictured in Fig. 8, but displaying the 2DR wave functions in terms of their loci of common nodes.}}
\end{figure}

\begin{figure}[h]
\center{\includegraphics[width=0.45\textwidth]{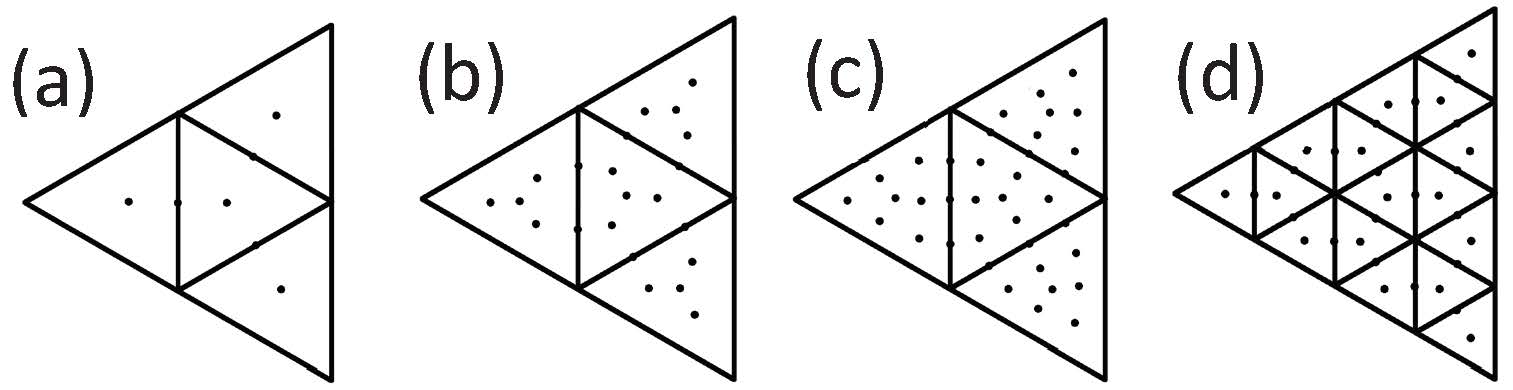}
\caption{Shown are some examples of higher index equilateral triangular box 2DR wave functions represented by both common points and equilateral triangles. (a) $\Psi^{(e,o)}_{2,4}$. (b) $\Psi^{(e,o)}_{2,6}$. (c) $\Psi^{(e,o)}_{4,6}$. (d) $\Psi^{(e,o)}_{4,8}$.}}
\end{figure}
We note that both with and without the equilateral triangular nodal lines inside the box, the loci of points and lines in each 2DR is invariant under all of the operations of $C_{3v}$.  Therefore, we classify those nodal loci as having symmetry $A_1$ in Table III.  We note that the common internal equilateral triangular nodal figures only arise for $(n,m\ne n+3p)$ both even.

\section{The equilateral triangular microstrip antenna}

\begin{table}
\vskip10pt
\begin{tabular}{lccccc}
\hline
\noalign{\vskip2pt}
Type&Symmetry&$|\Psi^{(e,o)}_{n,m}(x,y)\rangle$&$E$&$2R_3$&$3\sigma_v$\\
\noalign{\vskip3pt}
\hline
\noalign{\vskip3pt}
$A_1$&$x^2+y^2$&$|\Psi^{e}_{n,n+3p}(x,y)\rangle$&1&+1&+1\\
$A_2$&$y(3x^2-y^2)$&$|\Psi^{o}_{n,n+3p}(x,y)\rangle$&1&+1&-1\\
$A_1*$&fixed point nodes&$|\Psi^{(o,e,\theta)}_{n,m\ne n+3p}\rangle$ &2&-1&0\\
\noalign{\vskip3pt}
\hline
\noalign{\vskip3pt}
\end{tabular}
\caption{Representation types, symmetries, allowed 1DRs  $\Psi^{e,o}_{n,n+3p}(x,y)$ and 2DRs $\Psi^{(o,e)}_{n,m\ne n+3p}(x,y)$  of the thin equilateral triangular MSA, and operations of the $C_{3v}$ point group. For the 1DR wave functions, there  are 3 mirror planes $\sigma_v$ that bisect each angle, two rotations  $R_3$  by $\pm 2\pi/3$ about the centroid, and $E$ is the trace of the identity matrix for the appropriate group dimension.  For the 2DR wave functions, there are the same rotations, but only one $\sigma_v$  \cite{Tinkham}. *common nodal points.  See Fig. 13 and text.}
\end{table}
\begin{figure}[h]
\center{\includegraphics[width=0.48\textwidth]{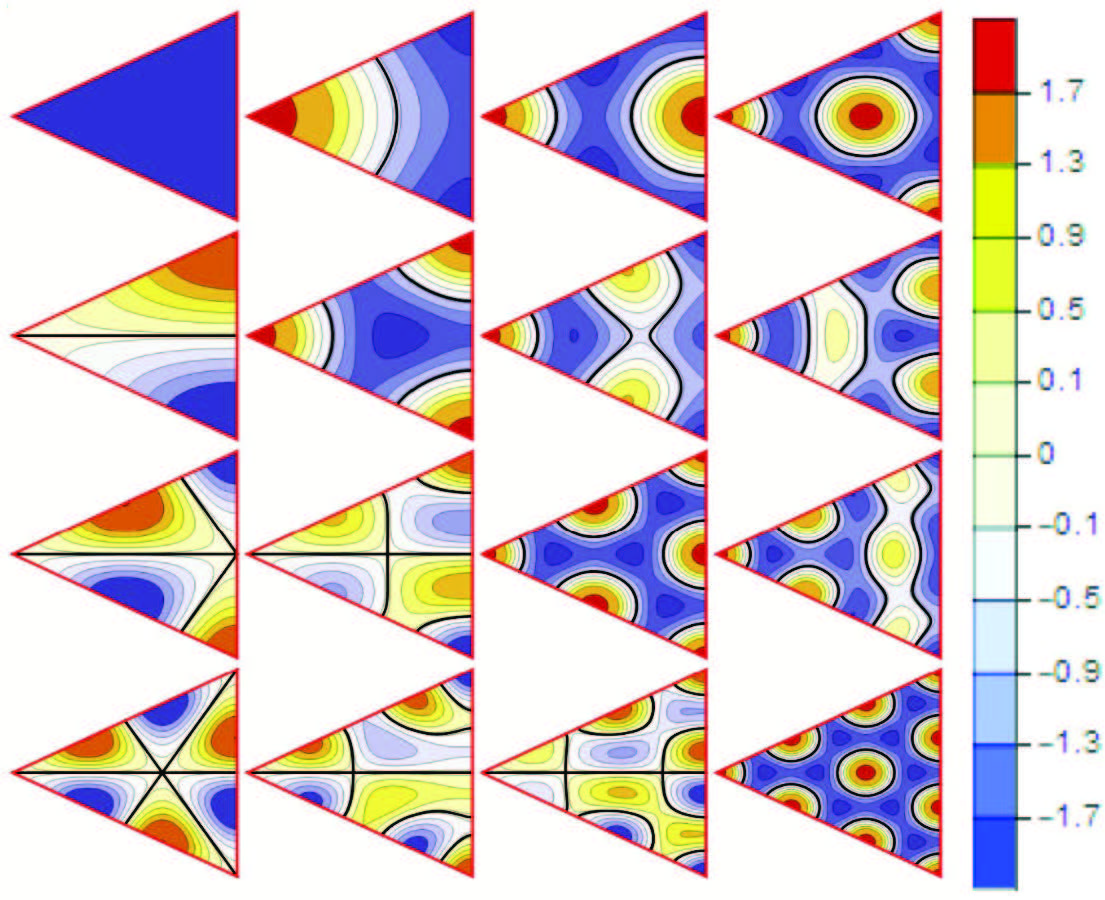}
\caption{(color online) Array of color-coded plots of the lowest frequency equilateral triangular microstrip antenna wave functions, generated from Eqs. (\ref{etmsaeven}), (\ref{etmsaodd}), and (\ref{msaconstants}), with $n,m=0,1,2,3$. The upper left solid blue figure has $f_{0,0}=0$, so it can be excluded.  The diagonal of the array represents  $\Psi^{e}_{n,n}(x,y)$, which for $n>0$ are 1DRs.  The $n\ne m$ even and odd wave functions lie respectively above and below the array  diagonal. The red boundaries indicate the Neumann conditions. The color code bar applies to each of these figures except the excluded (0,0) case.}}
\end{figure}

 Although the wave functions for the thin equilateral triangular MSA were given previously \cite{Cerkoney2017},  those authors only plotted the 1DR wave functions, and calculated the angular distributions of the output power from those resonant cavity modes and from the uniform Josephson current source at those mode frequencies.  Here we are primarily interested in contrasting the pictorial representation of the 1DR and 2DR wave functions.  We have

\begin{eqnarray}
\Psi^{e}_{\ell,m,n}(x,y)&\! =\!& A^e_{m,n}\!\biggl\{\cos\left[\left(\frac{2\pi x}{\sqrt{3}a}+\frac{2\pi}{3}\right)\ell\right]\nonumber\\
& &\times\cos\left[\frac{2\pi (m-n)y}{3a}\right]\>\>\>\>\>\> \nonumber\\
&& + \cos\left[\left(\frac{2\pi x}{\sqrt{3}a} + \frac{2\pi}{3}\right)m\right]\cos\left[\frac{2\pi (n-\ell)y}{3a}\right]\>\>\>\>\>\>\>\nonumber\\
&&+ \cos\left[\left(\frac{2\pi x}{\sqrt{3}a} + \frac{2\pi}{3}\right)n\right]\cos\left[\frac{2\pi (\ell-m)y}{3a}\right]\biggr\}\>\>\>\>\>\>\nonumber\\
& &\label{etmsaeven}
\end{eqnarray}
and
\begin{eqnarray}
\Psi^{o}_{\ell,m,n}(x,y)&\! =\!&A^o_{m,n}\biggl\{ \cos\left[\left(\frac{2\pi x}{\sqrt{3}a}+\frac{2\pi}{3}\right)\ell\right]\nonumber\\
& &\times\sin\left[\frac{2\pi (m-n)y}{3a}\right] \nonumber \\
&& + \cos\left[\left(\frac{2\pi x}{\sqrt{3}a} + \frac{2\pi}{3}\right)m\right]\sin\left[\frac{2\pi (n-\ell)y}{3a}\right]\nonumber\\
&&+ \cos\left[\left(\frac{2\pi x}{\sqrt{3}a} + \frac{2\pi}{3}\right)n\right]\sin\left[\frac{2\pi (\ell-m)y}{3a}\right]\biggr\}.\>\>\>\>\>\>\>\nonumber\\
& &\label{etmsaodd}
\end{eqnarray}
Each of the three terms for the even and odd wave functions must satisfy the EM wave equation ${\bm\nabla}^2\Psi+(k')^2\Psi=0$, as for the square MSA.  These forms can be shown to also obey the Neumann boundary conditions. As for the equilateral triangular box, we again have
\begin{eqnarray}
(n-m)(\ell+n+m)&=&0,
\end{eqnarray}
and the same arguments for $n=m$ in Section IV lead to the conclusion $\ell=-n-m$\cite{Cerkoney2017}.

The emission frequencies $f_{n,m}$  from an equilateral triangular MSA  with index of refraction $n_{\rm r}$ are then found to be
\begin{eqnarray}
f_{n,m}&=&\frac{2c_0}{3an_{\rm r}}\sqrt{m^2+n^2+mn}.
\end{eqnarray}

The
corresponding normalization coefficients are obtained by integrating $|\Psi^{o,e}_{\ell,m,n}|^2$ over the area of the equilateral triangle, and dividing by that area.  As was found previously \cite{Cerkoney2017},
\begin{eqnarray}
A_{m,n}^{e}& = &\left\{
        \begin{array}{ll}
            \frac{4}{3^{3/4}a}, & \quad m,n\ge 1, m\ne n \\
            \frac{2\sqrt{2}}{3^{3/4}a}, & \quad m>n=0, n>m=0,{\rm or}\hskip2pt m=n\hskip10pt,
        \end{array}
    \right.\nonumber\\
A_{m,n}^{o}& = &\left\{
                   \begin{array}{ll}
                   \frac{4}{3^{3/4}a}, &\quad   m,n\ge 1, m\ne n,\\
                   \frac{2\sqrt{2}}{3^{3/4}a}, &\quad  m>n=0,{\rm or}\hskip2pt n>m=0,\label{msaconstants}\\
                   0,&\quad m=n.\end{array}\right.
\end{eqnarray}
As shown in the previous section, these normalization constants are remarkably similar to those obtained for the equilateral triangular box.


\begin{figure}[h]
\center{\includegraphics[width=0.45\textwidth]{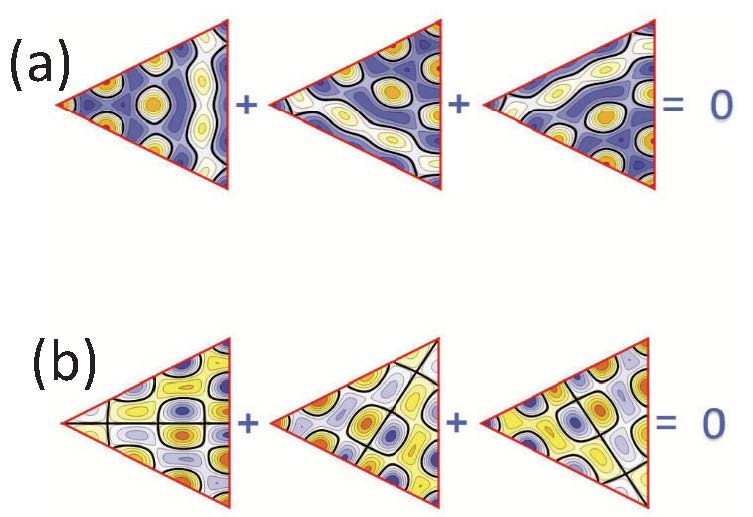}
\caption{(color online) (a) $|\Psi^{e}_{3,4}(x,y)\rangle$ and odd (b) $|\Psi^{o}_{3,4}(x,y)\rangle$ examples of 2DR equilateral triangular MSA wave functions  not pictured in Fig. 12 that each satisfy the picture equation $|\Psi\rangle +R_3|\Psi\rangle+R_3^2|\Psi\rangle=0$}}
\end{figure}

As for the 2DR wave functions for the equilateral triangular box that satisfy Eq. (\ref{2DREpsis}), the 2DR wave functions for the thin equilateral triangular MSA exhibit the same symmetries.  For example, in Fig. (13), we show pictorially that
\begin{eqnarray}
\!\!|\Psi^{(e,o)}_{3,4}(x,y)\rangle+R_3|\Psi^{(e,o)}_{3,4}(x,y)\rangle+R_3^2|\Psi^{(e,o)}_{3,4}(x,y)\rangle&\!=\!&0,\>\>\>\>\>\>\>
\end{eqnarray}
the only difference being the wave functions are not the box wave functions with Dirichlet boundary conditions given by Eqs. (12)-(16), but are instead given by Eqs. (24)-(28) for the MSA, which satisfy the Neumann boundary conditions with the normal derivative vanishing on each of the triangle's sides.  Therefore, the 2DR wave functions can be constructed exactly by analogy with Eqs. (18)-(21), also leading the analogous Nambu representation and the rank 2 matrices representing the identical symmetry operations $R_3$ and $\sigma_v^{(e,o)}$ given by Eqs. (22) and (23).  These thin equilateral MSA wave functions are also infinitely degenerate, as they also contain the arbitrary analogous mixing angle $\theta$.

\begin{figure}[!htb]
\center{\includegraphics[width=0.45\textwidth]{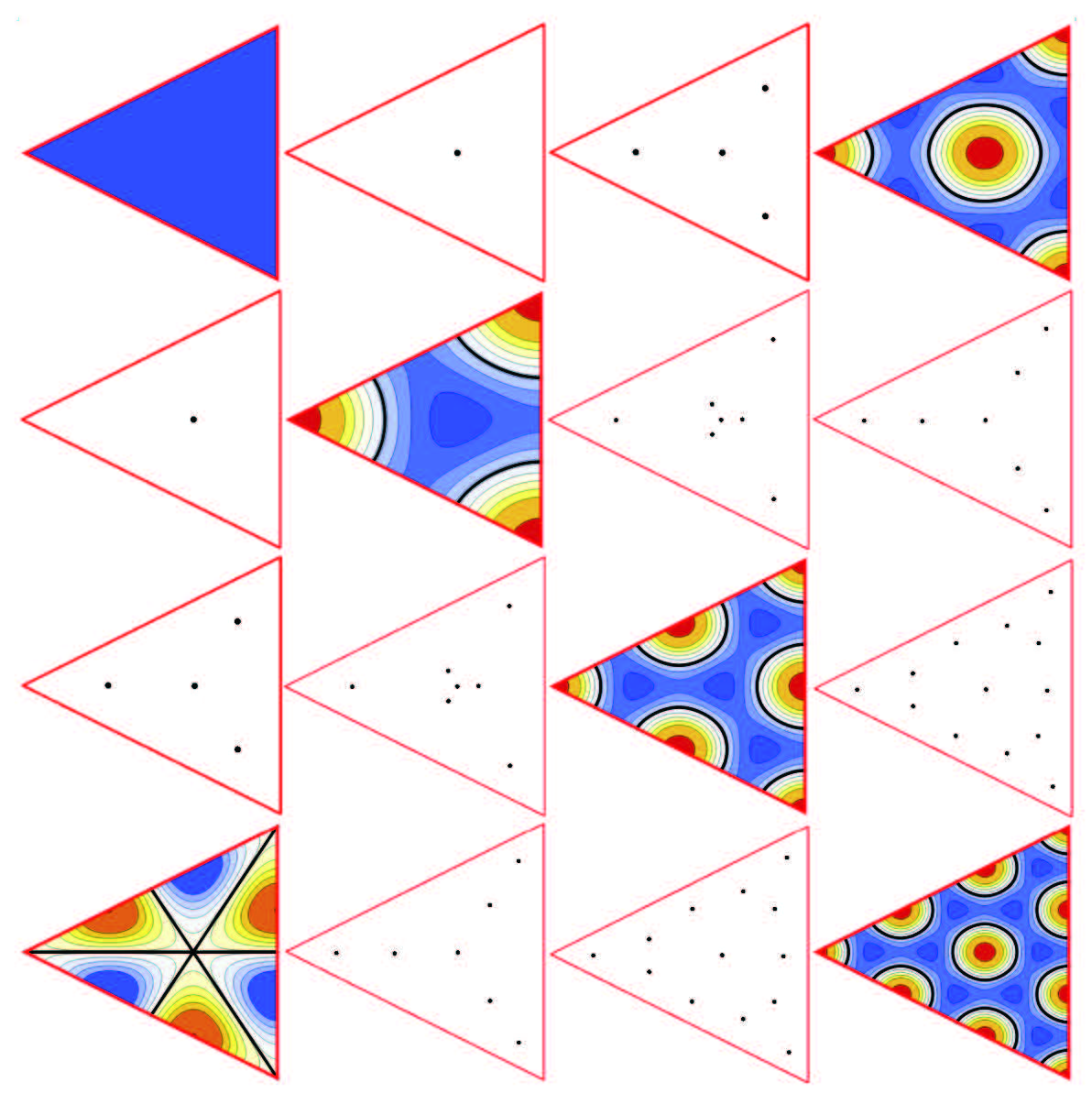}
\caption{(color online) Plots of the same equilateral triangular MSA wave functions pictured in Fig. 12, but displaying the 2DR wave functions in terms of their loci of common nodes.}}
\end{figure}
Thus, of the  equilateral triangular MSA wave functions pictured, those in color are 1DRs, and those along the array diagonal and in the upper right figure have $A_1$ symmetry, and the figure in the lower left has $A_2$ symmetry.  The rest of the figures are 2DRs,
and the pattern of nodal points has $A_1$ symmetry that is invariant under all operations of point group $C_{3v}$.  The matrices representing the 2$R_3$ and single $\sigma_v$ operations of the $C_{3v}$ point group acting upon the 2DR wavefunctions are identical to those described for the equilateral triangular box, except that the spatial parts of the wave functions are given by Eqs. (\ref{etmsaeven}) and (\ref{etmsaodd}).

We remark that the only difference between Tables III and IV for the equilateral triangular box and the thin equilateral triangular MSA is that some (with $m$ and $n$ both even) of the 2DR box wave functions contain both common nodal points and internal equilateral triangles (as well as on the boundary), but the 2DR MSA wave functions only contain common nodal points.  In both cases, the loci of the sets of nodal points and/or triangles are invariant under all operations of point group $C_{3v}$.

\section{The cylindrical box}

\begin{figure}[h]
\center{\includegraphics[width=0.48\textwidth]{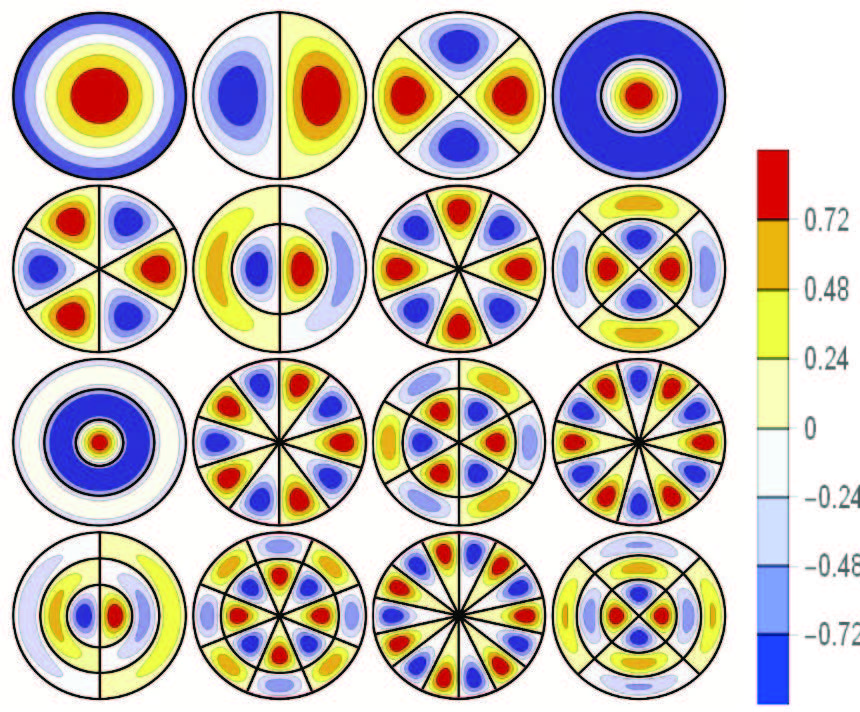}
\caption{(color online) Color coded plots of the 16 lowest energy cylindrical box wave functions, all oriented with $\theta=0$ and ranked in order from the top left to the bottom right array positions, listed as $(m,p)$:  Top row:  (0,1), (1,1), (2,1), (0,2).  Second row:  (3,1), (1,2), (4,1), (2,2).  Third row: (0,3), (5,1), (3,2), (6,1). Fourth row: (1,3), (4,2), (7,1), (2,3). Each figure has a qualitatively similar but numerically distinct color code bar. The color code bar shown is for the (2,2) mode.}} \end{figure}

For the quantum particle in a cylindrical box of radius $a$, the Schr{\"o}dinger equation is written in polar coordinates. $V(\rho)=0$ for $0\le\rho<a$, and $V(\rho)=\infty$ for $\rho\ge a$.  Using separation of variables and assuming $\Psi(\rho,\phi+2\pi)=\Psi(\rho,\phi)$, one obtains the Bessel equation with solutions of the first kind.  Since $\Psi(\rho,\phi)$ must be finite inside the cylindrical box, we only have the integer Bessel functions of the first kind, $J_m(k_m\rho)$ multiplied by $\sin(m\phi)$ or $\cos(m\phi)$.  Therefore a general state may be written
\begin{eqnarray}
\Psi_m(\rho,\phi)=[B_m\cos(m\phi)+C_m\sin(m\phi)]J_m(k_m\rho).\>\>\label{diskboxwavefunctions}
\end{eqnarray}
For the disk box of radius $a$, we require $\Psi_m(a,\phi)=0$, or
\begin{eqnarray}
J_m(k_ma)&=&0.
\end{eqnarray}
Since there are many possible zeroes of $J_m(x)$, we index them with $k_{m,p}$ values.  Thus, we set
\begin{eqnarray}
J_m(k_{m,p}a)&\!=\!&0,\\
\Psi_{m,p}(\rho,\phi)&\!=\!&[B_{m,p}\cos(m\phi)+C_{m,p}\sin(m\phi)]J_m(k_{m,p}\rho).\>\>\>\>\>\>\>\>\label{cylindricalwavefunctions}
\end{eqnarray}
It is immediately obvious that the cases $m=0$ and $m\ge1$ are qualitatively different. For $m=0$, the wave functions $\Psi_{0,p}(\rho)$ are 1DRs independent of $\phi$.  For $m\ge1$, the wave functions are all 2DRs. Since $\cos(m\phi)$ and $\sin(m\phi)$ are orthogonal when integrated over $\phi$ from 0 to $2\pi$, we could write either $B_{m,p}=A_{m,p}\cos(m\theta)$ and $C_{m,p}=A_{m,p}\sin(m\theta)$ or $B_{m,p}=-A_{m,p}\sin(m\theta)$ and $C_{m,p}=A_{m,p}\cos(m\theta)$.  Thus, we have two infinitely degenerate wave functions, which in Nambu form can be written as
\begin{eqnarray}
\Psi^{\theta}_{m,p}(\phi,\rho)&=&\Biggl(\begin{array}{c}\Psi^{(\theta,1)}_{m,p}(\phi,\rho)\\
\Psi^{(\theta,2)}_{m,p}(\phi,\rho)\end{array}\Biggr)\nonumber\\&=&A_{m,p}J_m(k_{m,p}\rho)\Biggl(\begin{array}{c}\cos[m(\phi-\theta)]\\ \sin[m(\phi-\theta)]\end{array}\Biggr),\>\>\label{diskbox2DR}
\end{eqnarray}
where $\theta$  satisfying $0\le\theta<2\pi$ is a 2DR wave function mixing angle as for the 2DR wave functions of the square and equilateral triangular boxes,
and the $A_{m,p}$ are found by normalization of both $|\Psi^{(1)}_{m,p}(\rho,\phi)|^2$ and $|\Psi^{(2)}_{m,p}(\rho,\phi)|^2$over the cross-sectional area of the cylinder,
\begin{eqnarray}
A_{m,p}&=&\left\{\begin{array}{lc}\frac{1}{a\sqrt{\pi\int_0^1xdxJ_m^2(\chi_{m,p}x)}},&m\ge1\\
\frac{1}{a\sqrt{2\pi\int_0^1xdxJ_0^2(\chi_{0,p}x)}},&m=0\end{array}\right.,\nonumber\\
\chi_{m,p}&=&k_{m,p}a.\label{cylindricalnormalizations}
\end{eqnarray}
\begin{figure}[h]
\center{\includegraphics[width=0.35\textwidth]{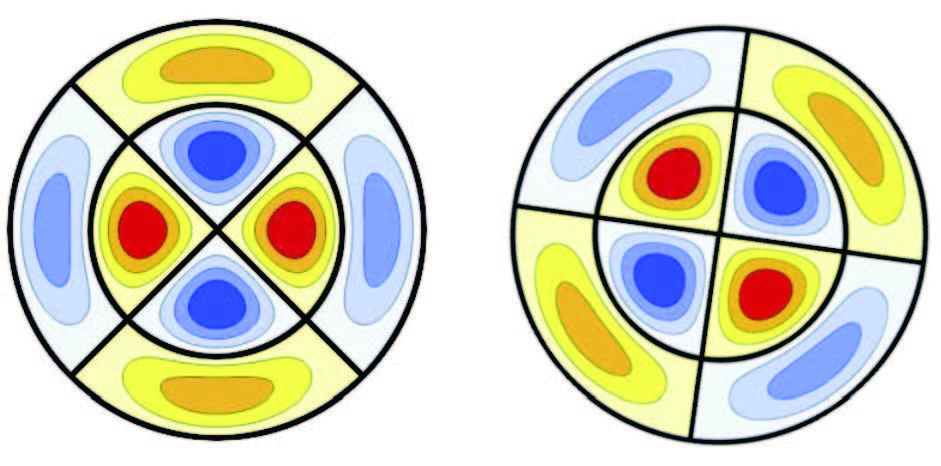}
\caption{(color online) Color-coded plots of $\Psi_{2,2}(\rho,\phi)$ with $\theta=0^{\circ}$ (left), and $\theta=53^{\circ}$ (right).}} \end{figure}
The energy of the $(m,p)$ mode for a spinless quantum particle of mass $M$ in the cylindrical box is given by
\begin{eqnarray}
E_{m,p}&=&\frac{\hbar^2\chi_{m,p}^2}{2Ma^2}.
\end{eqnarray}

\begin{figure}[h]
\center{\includegraphics[width=0.45\textwidth]{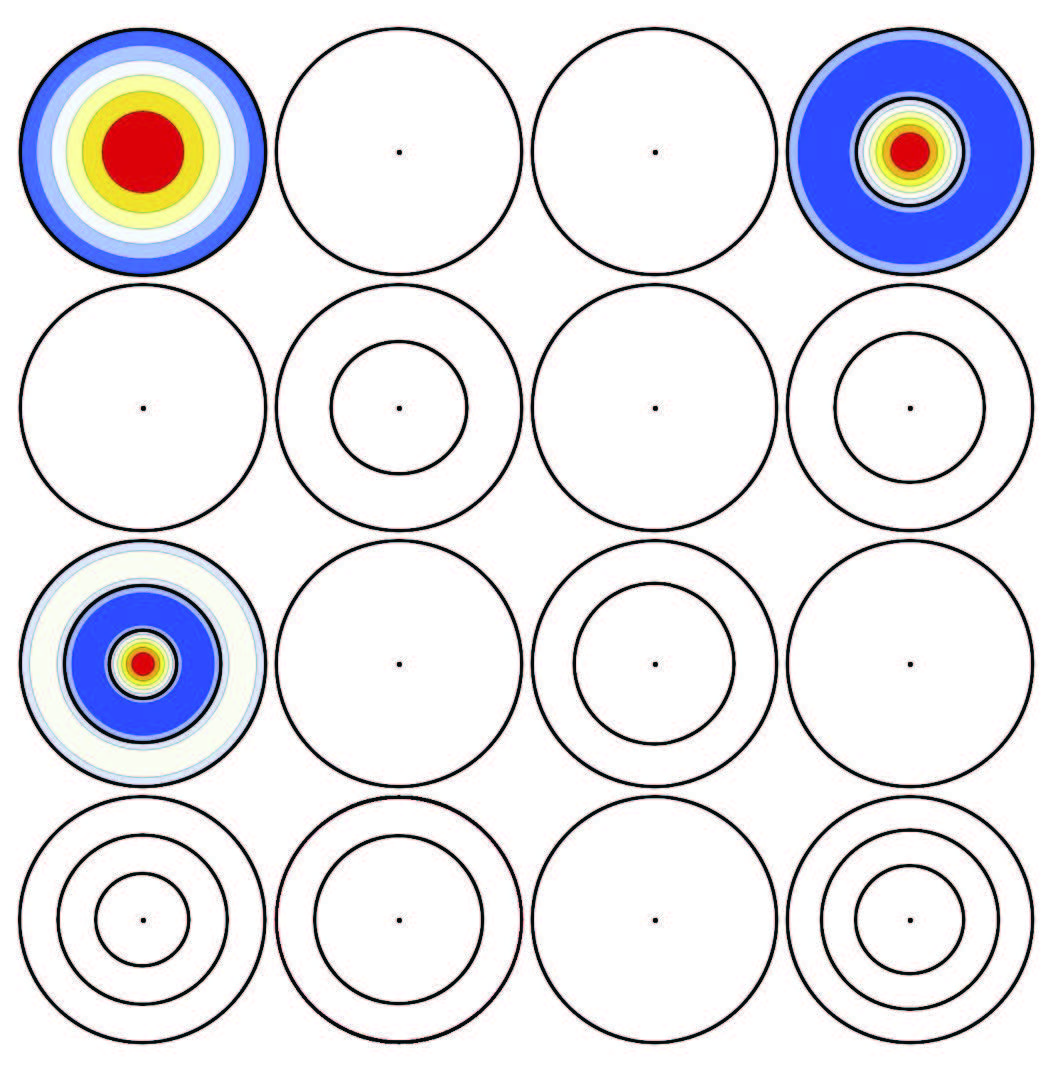}
\caption{Representing the cylindrical box wave functions in Fig. 15 to emphasize the differences between the non-degenerate $\Psi_{0,p}(\rho,\phi)$ and the infinitely degenerate $\Psi_{m,p}(\rho,\phi)$ for $m\ge1$.}}
 \end{figure}
A list of $\chi_{m,p}$ values for $0\le m\le 8$ and $1\le p\le 4$ is given in Table V.
\begin{table}
\vskip10pt
\begin{tabular}{lllll}
\hline
\noalign{\vskip2pt}
$m$&$\chi_{m,1}$&$\chi_{m,2}$&$\chi_{m,3}$&$\chi_{m,4}$\\
\noalign{\vskip3pt}
\hline
\noalign{\vskip3pt}
0&2.4048&5.52007&8.65372&11.79153\\
1&3.8317&7.01558&10.1734&13.32369\\
2&5.13562&8.41724 &11.61984&14.795981\\
3&6.38016&9.76102&13.01520&16.22346\\
4&7.58834&10.64709&14.3725&\\
5&8.77148&12.3386&15.70017&\\
6&9.936109&13.58929&17.0038&\\
7&11.08637&14.821268&&\\
8&12.22509&16.03777&&\\
\noalign{\vskip3pt}
\hline
\noalign{\vskip3pt}
\end{tabular}
\caption{Table of $\chi_{m,p}=k_{m,p}a$ for the cylindrical box of radius $a$. Missing table entries correspond to higher energy states than pictured in Fig. 15. }
\end{table}

In Fig. 15, we have presented an array of the 16 lowest energy wave functions for a quantum particle in a cylindrical box.  However, from Eq. (\ref{diskboxwavefunctions}), it is evident that the angle $\theta$ is arbitrary.  Since $0\le\theta<2\pi$, it can take on an infinite number of values, and hence cylindrical box eigenstates with $m\ne0$ are infinitely degenerate.  In Fig. 16, this arbitrary degeneracy is illustrated by comparing $\Psi_{2,2}(\rho,\phi)$ with $\theta=0$ and with its orientation with $\theta=53^{\circ}$.  Hence, those cylindrical  box wave functions with straight line nodes passing through their centroids are infinitely degenerate. Other than the circular line nodes at fixed $\rho$, the straight line nodes that can be rotated only have the single node at their common origin, the centroid.  This is illustrated in Fig. 17.

In order to construct the symmetry table, we first note that for $m\ge1$, Eq. (\ref{diskbox2DR}) contains two components, $\Psi^{(\theta,1)}_{m,p}(\phi,\rho)$ and $\Psi^{(\theta,2)}_{m,p}(\phi,\rho)$, which form the  orthonormal components of a 2DR wave function. When
 the two rotation matrices $R_{\pm m\varphi}$ for rotations by $\pm m\varphi$ about the $z$-axis normal to the centroid act on this Nambu form for the 2DR wave function, they  are easily found to be $R_{\pm m\varphi}={\bm 1}\cos(m\varphi) \pm i\sigma_y\sin(m\varphi)$, the traces of which are $2\cos(m\varphi)$.  This rotation matrix changes $\theta$ to $\theta\pm\varphi$ in the Nambu representation.  Since the bottom of the box is not a symmetry plane, there are no reflection planes.

  \begin{table}
\vskip10pt
\begin{tabular}{lccccc}
\hline
\noalign{\vskip2pt}
&&&circular&&\\
Type&Symmetry&$m$&nodes&E&$R_{\pm m\varphi}$\\
\noalign{\vskip3pt}
\hline
\noalign{\vskip3pt}
$A_1$&$x^2+y^2$&0&$p$&1&1\\
$A_1$*&centroid node&$\ge1$ &$p$&2&$2\cos(m\varphi)$\\
\noalign{\vskip3pt}
\hline
\noalign{\vskip3pt}
\end{tabular}
\caption{Representation types, symmetries,  and operations of the $C_{\infty v}$ point group for the cylindrical box. The number of circular nodes depends upon $p$.  $R_{\pm m\varphi}$ is a rotation about the centroid by the angle $\pm m\varphi$ and $E$ is the trace of the identity matrix for the appropriate group dimension \cite{Tinkham}. See Fig. 17 and text. *common nodal pattern.}
\end{table}

\begin{figure}[h]
\center{\includegraphics[width=0.48\textwidth]{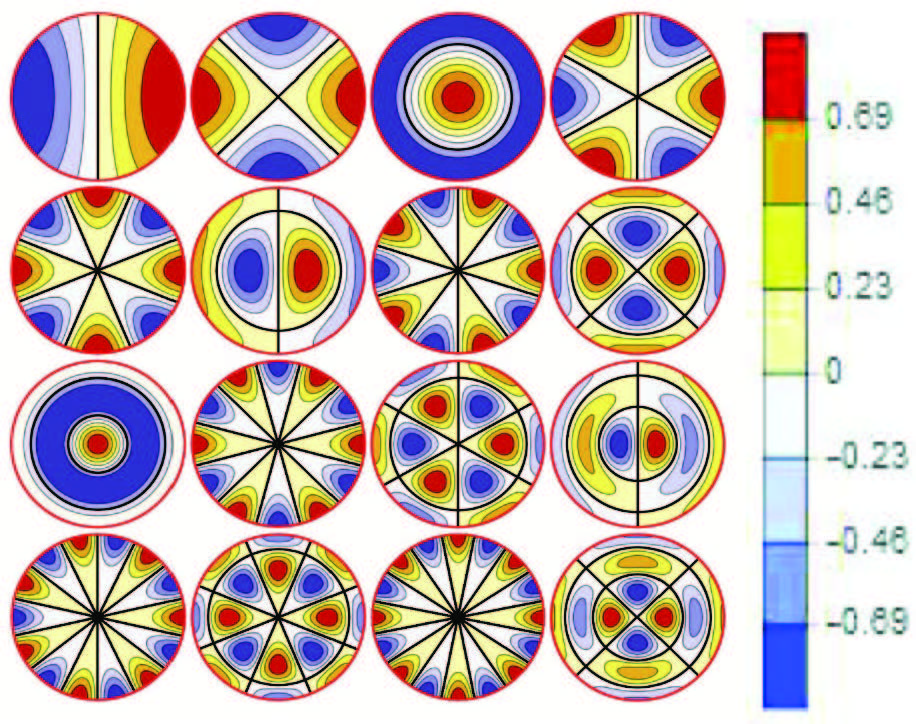}
\caption{(color online) Lowest frequency wave functions for the disk MSA \cite{Klemm2010b}. $f_{m,p}$ values increase from the top left to the bottom right. Top row, left to right:  (1,1), (2,1), (0,1), (3,1).  Second row: (4,1), (1,2), (5,1), (2,2).  Third row:  (0,2), (6,1), (3,2), (1,3).  Fourth row: (7,1),(4,2),(8,1),(2,3).  The red boundaries indicate the Neumann condition. The color code bars are qualitatively similar but numerically distinct for each of these figures. The color code shown is for the (2,1) mode.}}   \end{figure}

 \section{The disk microstrip antenna}
 The thin disk microstrip antenna has been studied previously \cite{Klemm2010b}.  Here we include it for two reasons: to compare the wave functions forms with those of the cylindrical box, and to use the degeneracy of the low-energy wave functions to correctly identify the experimentally measured resonant cavity mode emitted from a disk Bi2212 IJJ-THz emitter, which will be described in Section XIII.

 For the disk MSA, the wave functions also have the form of Eq. (\ref{diskboxwavefunctions}), but the boundary condition is different:
 \begin{eqnarray}
 \frac{dJ_m(k_m\rho)}{d\rho}\Bigr|_{\rho=a}&=&0.
 \end{eqnarray}
 As for the cylindrical box wave functions, there are an infinite number of such wave functions, which have the same forms as in Eq. (25), but the $\chi_{m,p}$ are different than those for the cylindrical box.  The amplitudes $A_{m,p}$ are also given by Eq. (\ref{cylindricalnormalizations}), but with the different $\chi_{m,p}$ values appropriate for the disk MSA.
 The emission frequencies $f_{m,p}$ from the 1DR cavity modes of the disk MSA are given by \cite{Klemm2010b}
 \begin{eqnarray}
 f_{m,p}&=&\frac{c_0\chi_{m,p}}{2\pi an_{\rm r}},
 \end{eqnarray}
 where the lowest group of $\chi_{m,p}$ values are listed in Table VII.
 \begin{table}
\vskip10pt
\begin{tabular}{lllll}
\hline
\noalign{\vskip2pt}
$m$&$\chi_{m,1}$&$\chi_{m,2}$&$\chi_{m,3}$&$\chi_{m,4}$\\
\noalign{\vskip3pt}
\hline
\noalign{\vskip3pt}
0&3.8317&7.0156&10.1735&13.3237\\
1&1.8412&5.3314&8.5363&11.7060\\
2&3.0542&6.7061 &9.9695&\\
3&4.2012&8.0152&11.3459&\\
4&5.3176&9.2824&&\\
5&6.4156&10.5199&&\\
6&7.5013&11.7349&&\\
7&8.5778&&&\\
8&9.6474&&&\\
9&10.7114&&&\\
10&11.7709&&&\\
\noalign{\vskip3pt}
\hline
\noalign{\vskip3pt}
\end{tabular}
\caption{Table of $\chi_{m,p}=k_{m,p}a$ for the disk microstrip antenna of radius $a$. Missing table entries correspond to higher energy states than pictured in Fig. 18. }
\end{table}

\begin{figure}[h]
\center{\includegraphics[width=0.35\textwidth]{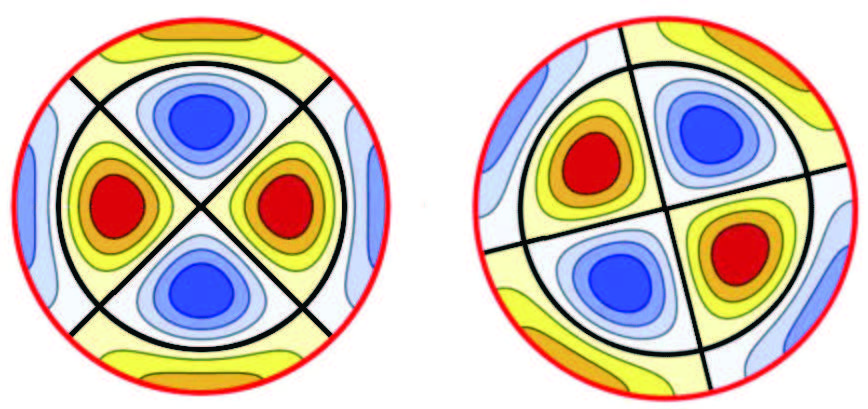}
\caption{(color online) Comparison of $\Psi_{2,2}(\rho,\phi)$ of the disk MSA with $\theta=0^{\circ}$ (left) and $\theta=31^{\circ}$ (right).}} \end{figure}
\begin{figure}[h]
\center{\includegraphics[width=0.45\textwidth]{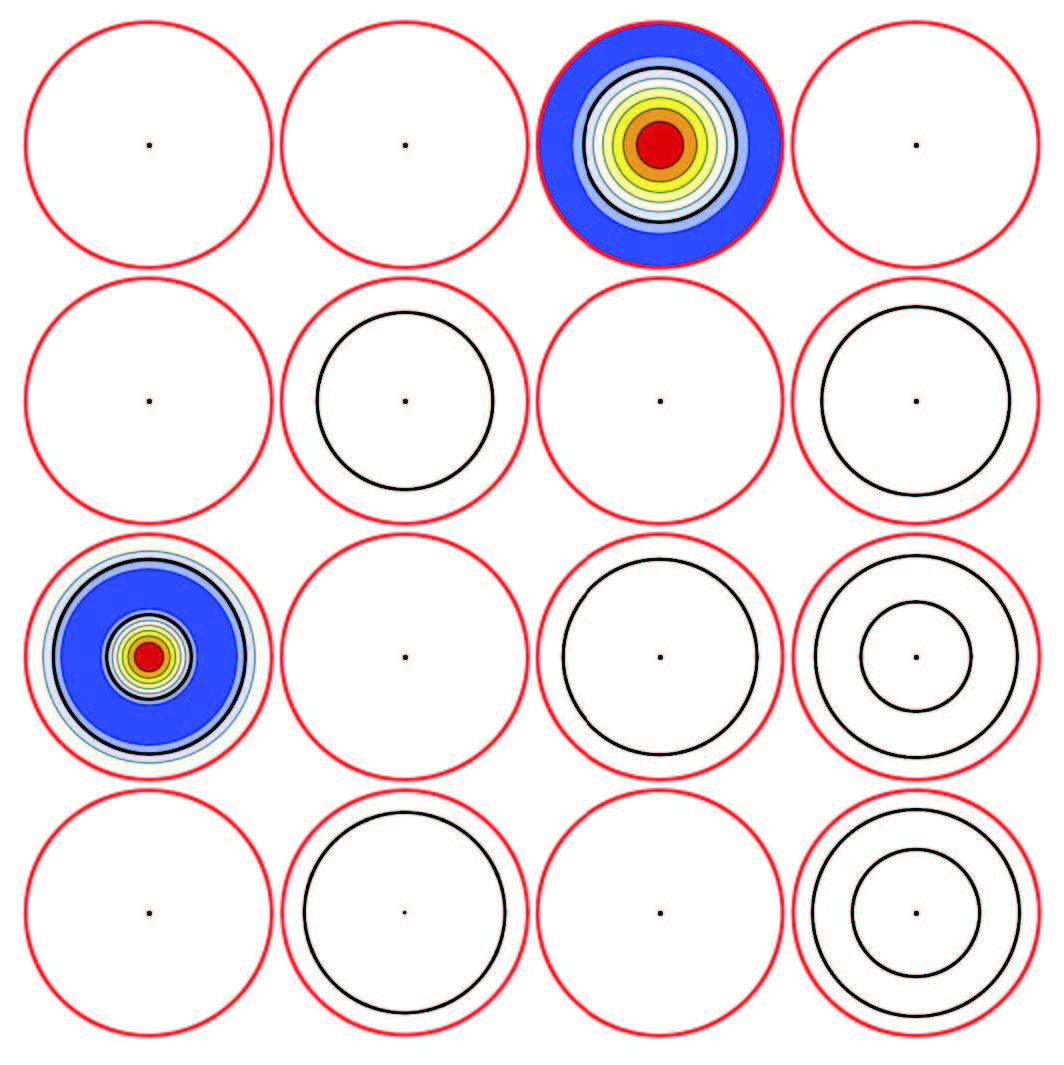}
\caption{(color online) Plots of the same disk MSA wave functions as in Fig. 18, but with the straight nodal lines passing though the centroid represented by a dot at the centroid for all of the 2DR wave functions.}}  \end{figure}

In conclusion, for the cylindrical  box and disk MSA, there are only two types of wave functions:  Non-degenerate wave functions with $m=0$ that have no nodal lines passing through the centroid, and a much larger class of infinitely degenerate wave functions with $m\ge1$ that have one or more nodal lines passing through the centroid.  When there is no reason to prefer one nodal line direction, the degeneracy is random, with an infinite number of possible directions.  The rotation matrices $R_{\pm m\varphi}$ for the 2DR MSA wave functions are identical to Eq. (\ref{diskbox2DR}) for the 2D cylindrical box, with the MSA $k_{m,p}$ values, the lowest energy values of which are given by Table VII. The only difference in symmetry Tables VI and VIII for the cylindrical box and the thin disk MSA is in the number of circular nodes for their 2DR wave functions, which is one more for the box due to the boundary condition.  Thus, we conclude that for the thin MSAs, the only modes that can build up a cavity resonance are those of the 1DR wave functions with $m=0$.
 \begin{table}
\vskip10pt
\begin{tabular}{lccccc}
\hline
\noalign{\vskip2pt}
&&&circular&&\\
Type&Symmetry&$m$&nodes&E&$R_{\pm m\varphi}$\\
\noalign{\vskip3pt}
\hline
\noalign{\vskip3pt}
$A_1$&$x^2+y^2$&0&$p$&1&1\\
$A_1$*&centroid node&$\ge1$ &$p-1$&2&$2\cos(m\varphi)$\\
\noalign{\vskip3pt}
\hline
\noalign{\vskip3pt}
\end{tabular}
\caption{Representation types, symmetries, and operations of the $C_{\infty v}$ point group for the disk MSA wavefunctions $\Psi_{m,p}(\phi,\rho)$.   The number of circular nodes depends upon $p$.  The  $R_{\pm m\varphi}$ are rotations about the $z$ axis normal to the centroid by $\pm m\varphi$ and $E$ is the trace of the identity matrix for the appropriate group dimension \cite{Tinkham}. See Fig. 20 and text. *common nodal pattern.}
\end{table}

\begin{figure}[t]
\center{\includegraphics[width=0.45\textwidth]{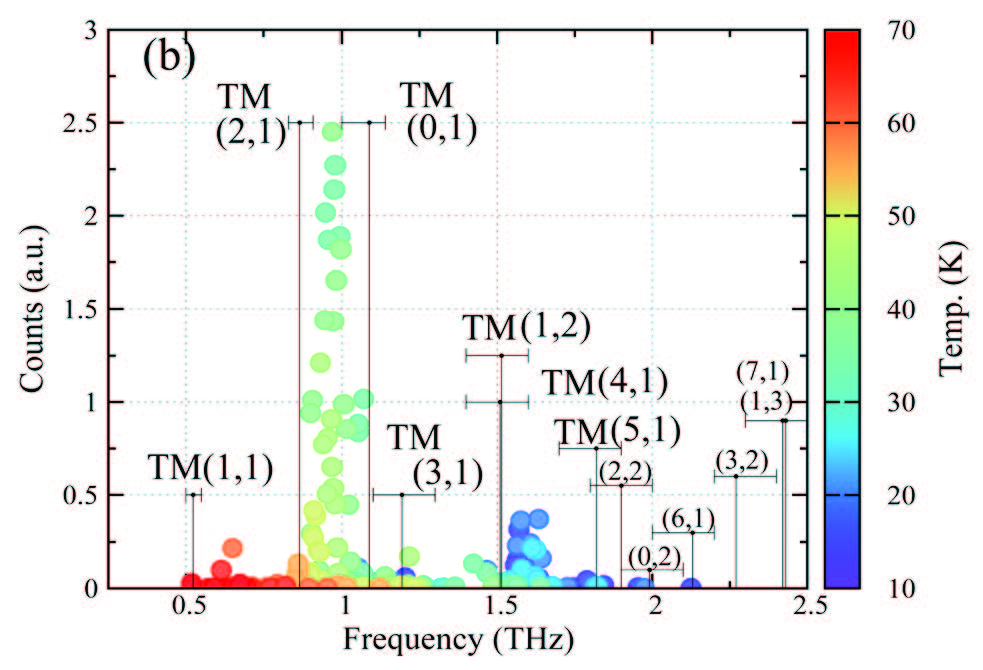}
\caption{Frequency dependence of the emission from a stand-alone Bi2212 disk mesa \cite{Kashiwagi2015}. Reprinted with permission of  T. Kashiwagi, K. Sakamoto, H. Kubo, Y. Shibano, T. Enomoto, T. Kitamura, K. Asanuma, T. Yasui, C. Watanabe, K. Nakade, Y. Saiwai, T. Katsuragawa, M. Tsujimoto, R. Yoshizaki, T. Yamamoto, H. Minami, R. A. Klemm, and K. Kadowaki, A high-$T_c$ intrinsic Josephson junction emitter tunable from 0.5 to 2.4 terahertz, Appl. Phys. Lett. {\bf 107}, 082601 (2015). \copyright 2015 AIP Publishing LLC.}} \end{figure}
\begin{figure}[h]
\center{\includegraphics[width=0.45\textwidth]{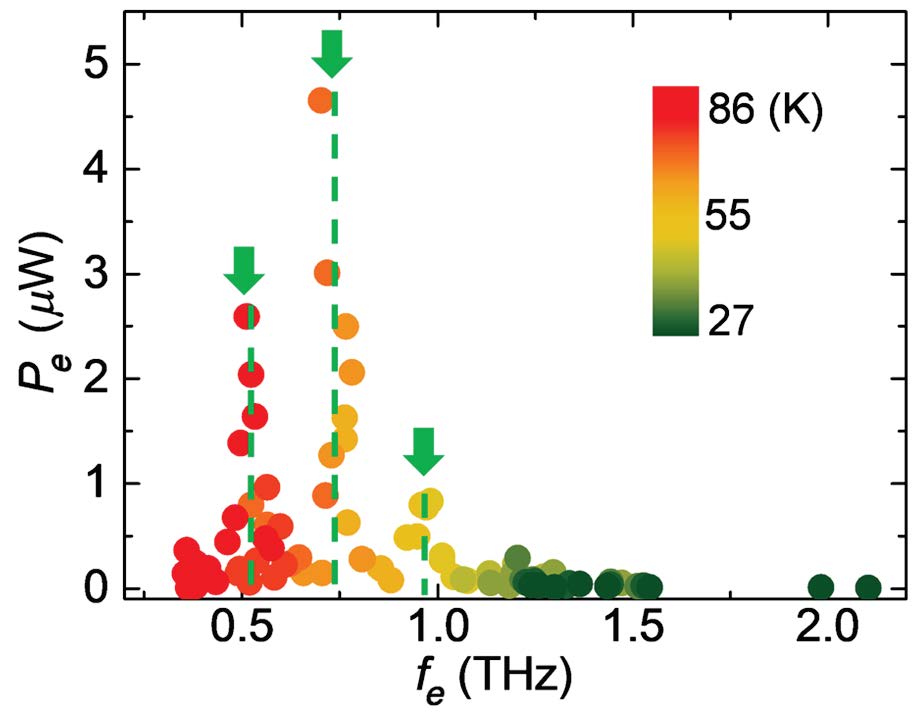}
\caption{Frequency dependence of the emission from a square stand-alone Bi2212 mesa \cite{Sun2018}. Reprinted with permission of  H. Sun, R. Wieland, Z. Xu, Z. Qi, Y. Lv, Y. Huang, H. Zhang, X. Zhou, J. Li, Y. Wang, F. Rudau, J. S. Hampp, D. Koelle, S. Ishida, H. Eisaki, Y. Yoshida, B. Jin, V. P. Koshelets, R. Kleiner, H. Wang, and P. Wu, Compact high-$T_c$ superconducting terahertz emitter operating up to 86 K, Phys. Rev. Appl. {\bf 10}, 024041 (2018). \copyright 2018 American Physical Society. }}
 \end{figure}

\section{Comparison with experiments}

Since the original discovery of coherent THz emission from the IJJs in Bi2212 \cite{Ozyuzer2007}, a variety of experimental groups in many countries have been working on the topic, trying to understand its properties and to increase the output power.  There have so far been six review articles on the subject \cite{Kashiwagi2012,Welp2013,Kakeya2016,Kashiwagi2017,Kleiner2019,Delfanazari2020}.  In the early work, the first type of Bi2212 mesas were formed by subjecting a cleaved single crystal of Bi2212 to an Ar beam with a mask, that cut into the unmasked region of the crystal, leaving a standing mesa with the remainder of the Bi2212 crystal as the substrate.  A second type was a groove mesa, obtained by simply cutting a groove into the top of a cleaved Bi2212 crystal, which was first done for  groove rectangular, square, and disk mesas \cite{Tsujimoto2010}, and subsequently for  a groove equilateral triangular mesa \cite{Delfanazari2013}. In those experiments, the emission frequencies for 3 groove disk mesas, the rectangular, the square, and 3 groove equilateral triangular mesas were all consistent with their respectively lowest-frequency TM(1,1) and TM(0,1) modes \cite{Tsujimoto2010,Delfanazari2013}.   For the rectangular mesa, the TM(0,1) mode with a nodal line bisecting the mesa length  is non-degenerate, and as expected, was the first shape to be shown to build up a cavity resonance \cite{Ozyuzer2007}.   However, as discussed in Sections II, V,  and VII, those lowest frequency TM(0,1) square and equilateral triangular modes and the lowest frequency TM(1,1) disk modes are all infinitely degenerate, and if the groove mesas were sufficiently accurately constructed for those symmetries to be relevant, shouldn't build up a cavity resonance at those frequencies \cite{Klemm2010a,Klemm2010b}.  The facts that the lowest frequency, infinitely-degenerate cavity modes were observed in those experiments were most likely due to the breaking of the square, equilateral triangular, or disk symmetry due to some phenomenon that was not understood at the time of those early experiments.

The problem turned out to be that the introduction of a dc $V$ and current $I$ across the stack of IJJs in Bi2212 leads to severe Joule heating effects, resulting in hot spots, spatial regions in which $T > T_c$ \cite{Wang2009,Wang2010,Minami2014,Kashiwagi2017b}.  For a rectangular Bi2212 mesa, these hot spots were  observed by laser scanning \cite{Wang2009,Wang2010}, SiC photoluminescence \cite{Minami2014}, and thermoreflectance microscopy \cite{Kashiwagi2017b}, and when a hot spot develops away from the center of a square, equilateral triangular, or disk mesa, it breaks the symmetry, and allows for the emission of photons at the lowest-frequency from the infinitely-degenerate respective modes.  But it was suggested that removing the mesa from its superconducting substrate and coating the top and bottom with a perfect electric conductor such as Au, the output power could be enhanced \cite{Klemm2010a,Klemm2010b}. After doubly cleaving a Bi2212 sample from a single crystal, the top and bottom surfaces are first covered with about 50-100 nm of Ag, followed by about 50-100 nm of Au. These mesas with thin Au layers on their top and bottom surfaces are presently known as either as ``stand-alone'' mesas \cite{Kashiwagi2015b}, or  as ``GBG'' for ``gold-Bi2212-gold'' mesas \cite{Sun2018}. Since Au is a superior thermal conductor, as long as the stand-alone mesas are not much thicker than 1-2 $\mu$m, it is usually possible to avoid most of the heating problems, including the development of hot spots.  An efficient procedure to manufacture the stand-alone Bi2212 mesas was published \cite{Kashiwagi2015b}. In such mesas, the amount of Joule heating was greatly reduced, and it became possible to investigate experimentally the effects of mesa symmetry upon the cavity resonances observed.

In Fig. 21, the frequency dependence of the emission from a Bi2212 stand-alone disk mesa is shown. Unlike the emission data from three groove disk mesas studied earlier \cite{Tsujimoto2010}, the stand-alone disk mesa did not build up a cavity resonance at the infinitely degenerate TM(1,1) disk MSA mode.  Instead, a strong cavity resonance appeared at 1.0 THz, in-between the predicted resonance frequencies of the TM(0,1) and TM(2,1) modes.   However, since the stand-alone Bi2212 disk mesa was sandwiched in-between two sapphire substrates, the substrates might cause a slight shift in the cavity resonance frequency.  Although we are not aware of calculations for a MSA sandwiched between two substrates, when a MSA is sitting atop a sapphire substrate, there is a slight downward shift in the emission frequency \cite{Balanis}.  In addition, since the infinitely degenerate TM(1,1) mode was not excited, it seems reasonable to assume that the infinitely degenerate TM(2,1) mode would also not be excited.  For those two reasons, we assign the strong emission at 1 THz to the non-degenerate TM(0,1) disk MSA mode.  From the data, it appears that the downward frequency shift due to the two sapphire substrates is approximately 3\% for emission at 1.0 THz.

More recently, the emissions from a stand-alone square Bi2212 mesa with $a=200 \mu$m  sandwiched between sapphire substrates was studied \cite{Sun2018}, and the frequency dependence of the emission from that mesa is shown in Fig. 22.  From Eq. (11), it is possible to analyze the emission spectrum in terms of the possible mode frequencies. We note that the authors originally misidentified the cavity resonance mode indices, but corrected them in an erratum \cite{Sun2018}. A table of mode indices, the degeneracies, and calculated frequencies  without and with a 3\% substrate effect is given in Table IX.  We note that the lowest two frequencies are at or below the low-temperature Josephson plasma frequency $f_p\approx 0.250$ THz \cite{Singley2004}, and are screened out by the Josephson plasma.

 \begin{table}
\vskip10pt
\begin{tabular}{lccc}
\hline
\noalign{\vskip2pt}
$(n,m)$&$g$&$f_{n,m}$ (THz)&$f'_{n,m}$ (THz)\\
\noalign{\vskip3pt}
\hline
\noalign{\vskip3pt}
(1,0)&$\infty$&0.179*&0.174*\\
(1,1)&1&0.253*&0.245*\\
(2,0)&2&0.357&0.346\\
(1,2)&$\infty$&0.399&0.387\\
(2,2)&1&0.505&0.490\\
(3,0)&$\infty$&0.536&0.520\\
(3,1)&2&0.564&0.547\\
(3,2)&$\infty$&0.644&0.625\\
(4,0)&2&0.714&0.693\\
(4,1)&$\infty$&0.736&0.714\\
(3,3)&1&0.758&0.735\\
(4,2)&2&0.799&0.775\\
(4,3)&$\infty$&0.892&0.865\\
(5,0)&$\infty$&0.892&0.865\\
(5,1)&2&0.910&0.883\\
(5,2)&$\infty$&0.962&0.933\\
(4,4)&1&1.010&0.980\\
(5,3)&2&1.04&1.01\\
\noalign{\vskip3pt}
\hline
\noalign{\vskip3pt}
\end{tabular}
\caption{Predicted cavity mode frequencies $f_{n,m}$ from Eq. (10), the estimated $f'_{n,m}=0.97f_{n,m}$ due to the substrate in THz, the degeneracy $g$ of the $(n,m)$ mode for a square stand-alone Bi2212 mesa of side $200 \mu$m and $n_{\rm r}=4.2$. *At or below the Josephson plasma frequency $f_p\approx0.250$ THz \cite{Singley2004}.}
\end{table}

It is not clear that one should assume the substrate reduction factor to be the same percentage for each frequency measured.  But the resonances near to 1.0 THz are likely to have nearly the same shift in the two experiments. Hence a strong case can be made that the resonances are mostly associated with the non-degenerate $(n,n)$ modes. It is certainly true that the doubly degenerate $(n,n+2p)$ modes have orthogonal 1DR symmetries, so on some time scale, it would be difficult for the system to stick with one symmetry and to ignore the other one. That is, if the system oscillates between the two symmetries on a time scale inverse to the common mode frequency, there would be no cavity resonance.  The data are consistent with this notion \cite{Sun2018}.
\section{Summary and conclusions}
We have studied the wave functions of high-symmetry two-dimensional quantum boxes and electromagnetic microstrip antennas (or cavities).  The symmetries studied are those of a square, an equilateral triangle, and a disk. Each of these symmetries has one- and two-dimensional representations of its wave functions. The two-dimensional representations are infinitely degenerate \cite{Tinkham}.  In addition, for square boxes and microstrip antennas, there are doubly-degenerate cases, the wave functions of which can be written in terms of orthogonal one-dimensional representations.

Although the two-dimensional quantum box might have some approximate experimental relevance to quantum wells, the main interest from the experimental side is for thin microstrip antennas.  This is particularly of importance for the coherent THz emission from the intrinsic Josephson junctions in the layered high-temperature superconductor, Bi$_2$Sr$_2$CaCu$_2$O$_{8+\delta}$, or Bi2212.  In the early days of the coherent THz emission from Bi2212, there were severe heating effects that interfered with accurate comparisons of theory with experiment.  Now that thermally-managed stand-alone Bi2212 (or Au-Bi2212-Au) mesas are the primary devices under study, the effect of the degeneracies of the modes is important to consider.  The comparison of the experimental output from stand-alone Bi2212 disk and square mesas provide experimental evidence that the non-degenerate modes are the ones that can be excited in order to increase the output power.  This suggests the construction of stand-alone mesas that have only or predominantly non-degenerate modes.

\begin{figure}
\center{\includegraphics[width=0.48\textwidth]{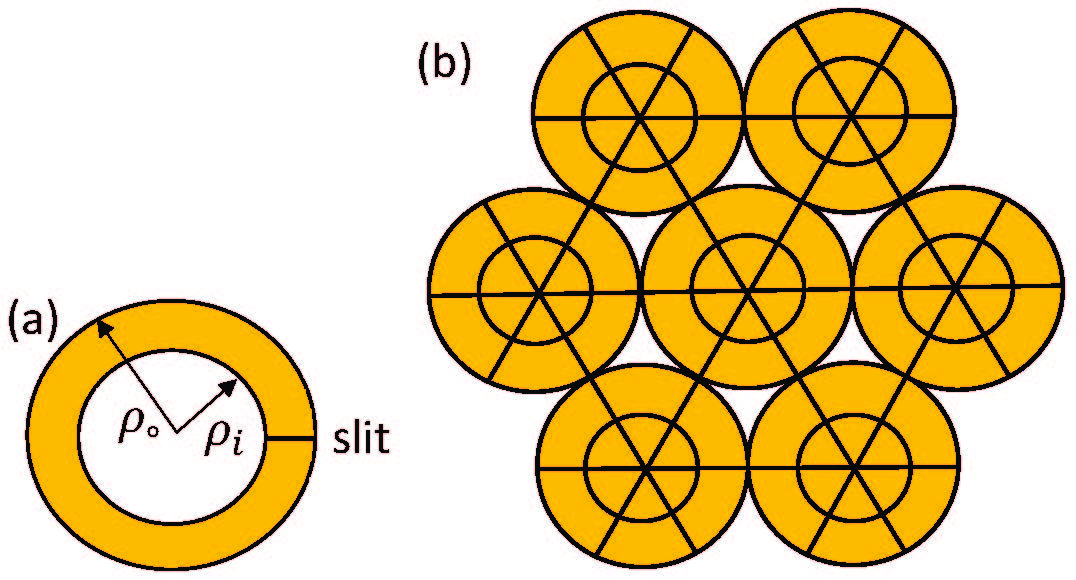}
\caption{(a) Sketch of an annular stand-alone mesa with a single slit. (b) Sketch of an array of 84 stand-alone mesas.  Seven stand-alone disk mesas are fixed in a hexagonal close packed array, and are equally cut circularly with an atomic beam into smaller disks and  annulli. Then, with straight line slits, the original disks are each cut into 12 mesas, and the entire array of 84 mesas could be close enough to one another to emit coherently. \cite{Delfanazari2020}}}
\end{figure}

The simplest example is that of a rectangle in which the ratio of length to width is not that of two integers, stand-alone mesas of which showed excitations at many frequencies \cite{Kashiwagi2018}.   A singly-slitted annulus  has been suggested as another possibility \cite{Delfanazari2020}, and our independent analysis has shown that the modes odd and even about the slit are not degenerate \cite{Shouk2021}.  But another possibility is a disk that is cut into 12 pieces of two different types:  The disk is first cut into a smaller disk and an annulus by a He or Ar beam, and those two objects are cut with three straight cuts at angles 60$^{\circ}$ apart, dividing the  smaller disk and annulus into six equivalent pie-shaped wedge mesas and six equivalent hexaslitted annuli.  These would all fit closely together, allowing for the possibility of coherent emission from a much larger number of intrinsic Josephson junctions, increasing the output power well above 1 mW.  Our preliminary studies showed that all of the modes are nondegenerate.

 Such a device has been proposed recently  \cite{Delfanazari2020}, and is redrawn in Fig. 23.  In order to maximize the probability of matching the resonant frequencies of the inner pie-shaped wedge mesas with the outer hexaslitted annular mesas, the ratio of $\rho_i/\rho_o$ is varied, and the resonant frequencies can be calculated to find if at least two frequencies from both shapes will match in the 1-2 THz range for which the output power of compact continuous wave coherent sources is generally less than the 1 mW necessary for practical applications \cite{Delfanazari2020}.

Since the only equilateral triangular microstrip antennas of Bi2212 were made prior to the construction of stand-alone mesas \cite{Delfanazari2013,Delfanazari2013b}, they most likely exhibited hot spots, breaking the symmetry and allowing the infinitely-degenerate TM(0,1) mode to be observed.  We therefore encourage experiments on stand-alone equilateral triangular Bi2212 mesas to compare with our predictions, as was done with stand-alone disk and square mesas \cite{Kashiwagi2015,Sun2018}.
\section{Acknowledgments}
The authors acknowledge discussions with Kazuo Kadowaki and Allan H. MacDonald. R. A. K. was supported in part by the U. S. Air Force Office of Scientific Research (AFOSR) LRIR \#18RQCOR100, and the AFRL/SFFP Summer Faculty Fellowship Program provided by AFRL/RQ at WPAFB.

J. R. R., P. Y. C., and A. B. contributed equally to this work.  R. A. K. supervised the work and wrote the paper.

\end{document}